\documentclass[final,3p,times,authoryear]{elsarticle}

\usepackage[T1]{fontenc}
\usepackage[utf8]{inputenc}
\usepackage{amsmath,amssymb,amsthm}
\usepackage{graphicx}
\usepackage{xcolor}
\usepackage{booktabs}
\usepackage{array}
\usepackage{longtable}
\usepackage{enumitem}
\usepackage{caption}
\usepackage{subcaption}
\usepackage{mathtools}
\usepackage{bm}
\usepackage{hyperref}
\hypersetup{colorlinks=true,
            linkcolor=blue!60!black,
            citecolor=green!50!black,
            urlcolor=blue!60!black}

\newcommand{\Msun}{M_{\odot}}
\newcommand{\kpc}{\,\mathrm{kpc}}
\newcommand{\kms}{\,\mathrm{km\,s^{-1}}}
\newcommand{\vc}{v_{c}}
\newcommand{\Rsun}{R_{\odot}}
\newcommand{\Rvir}{R_{\mathrm{vir}}}
\newcommand{\Mvir}{M_{\mathrm{vir}}}
\newcommand{\rholocDM}{\rho_{\mathrm{DM},\odot}}

\newcommand{\ie}{i.e.}
\newcommand{\dif}{\mathrm{d}}
\newcommand{\pp}{\partial}

\theoremstyle{definition}

\journal{New Astronomy Reviews}

\begin{document}

\begin{frontmatter}

\title{The Rotation Curve of the Milky Way: State of the Art,
       the Keplerian Decline Debate, and Implications for Dark Matter}

\author[roma1]{Alessandro Melchiorri\corref{cor1}}
\ead{alessandro.melchiorri@uniroma1.it}

\cortext[cor1]{Corresponding author}

\address[roma1]{Dipartimento di Fisica ``G.\ Marconi'',
  Sapienza Universit\`{a} di Roma, Piazzale Aldo Moro 2, I-00185 Roma, Italy.
  INFN, Sezione di Roma 1, Roma, Italy.}

\author[spain]{Ruchika}
\ead{ruchika.science@usal.es}

\address[spain]{Departamento de Física Fundamental and IUFFyM, Universidad de Salamanca, E-37008 Salamanca, Spain}

\begin{abstract}
For four decades, the roughly flat rotation curve (RC) of the Milky Way
(MW) stood as local evidence for an extended dark matter (DM) halo.
\textit{Gaia} DR3 has changed the picture: several analyses now
find a velocity decline beyond $R \approx 15\kpc$, and the most
radical interpretation \citep{jiao2023} claims a nearly Keplerian fall-off
($\vc \propto R^{-1/2}$) that excludes a flat RC at $3\sigma$ and implies a
total dynamical mass of only $\sim 2 \times 10^{11}\Msun$---three to five
times below pre-\textit{Gaia} estimates.  Taken at face value, this would
make the MW exceptional among comparable spirals and challenge both
$\Lambda$CDM and MOND.  It would also alter predictions for direct-detection
experiments, although the local DM density is not fixed by virial mass
alone.  The Keplerian claim rests on
the delicate assumptions of axisymmetric Jeans modelling, while stellar
streams, globular clusters, satellite kinematics, and the Local Group
timing argument generally favour a substantially heavier halo, and cosmological
simulations reveal potentially substantial biases in Jeans-inferred
outer RCs.   This review offers a
self-contained, pedagogical account of the
debate: we derive the full formalism from first principles---the Jeans
equations, the asymmetric drift correction, the standard DM halo and
baryonic mass models, the MOND flat-RC prediction, and the timing
argument---present an illustrative phenomenological MCMC fit and a
Gaussian Process reconstruction of the RC, and critically
assess which features of the decline can be regarded as established and
which remain open.
\end{abstract}


\end{frontmatter}

\tableofcontents

\section{Introduction}
\label{sec:intro}

\subsection{The rotation curve as a mass probe}

The rotation curves of spiral galaxies are among the most consequential
measurements in modern astrophysics: what began as an attempt to weigh
galaxies became, within roughly a decade, the most widely cited dynamical
evidence for dark matter.  The Milky Way occupies an awkward position in
that history.  It is the only galaxy whose individual stars can be
observed in full six-dimensional phase space, and at the same time the one
for which the rotation curve is hardest to extract, precisely because we
observe it from within the disc we are trying to measure.  Whether its
outer rotation curve is flat, gently declining, or Keplerian is the
question this review addresses; we begin with the elementary relation
between circular speed and enclosed mass that underlies the entire
discussion.

The circular speed $\vc(R)$ of a test particle orbiting at Galactocentric
distance $R$ in the equatorial plane of a galaxy is determined by the total
gravitational potential $\Phi(R)$ through
\begin{equation}
  \vc^2(R) = R\left.\frac{\pp\Phi}{\pp R}\right|_{z=0},
  \label{eq:vc_basic}
\end{equation}
where $z$ is the height above the midplane.  For a spherically symmetric
mass distribution this becomes
\begin{equation}
  \vc^2(r)=\frac{G M(<r)}{r}.
  \label{eq:vc_spherical_mass}
\end{equation}
Equation~\eqref{eq:vc_spherical_mass} is not an exact enclosed-mass relation
for a flattened disc.  For a general axisymmetric disc--halo system,
$M_{\rm sph}(<R)\equiv R\vc^2/G$ is therefore a
\textit{spherical-equivalent} dynamical mass; recovering the true enclosed
mass requires a model for the three-dimensional mass geometry.

For an isolated point mass $M$, Newton's law gives $\vc = \sqrt{GM/R}$,
the Keplerian velocity that decreases as $R^{-1/2}$.  The same behaviour would
be expected in a galaxy if most of the mass were concentrated within the
optical disc: once $R$ exceeds the radius enclosing essentially all the mass,
$\vc$ should fall as $R^{-1/2}$.  This is indeed observed in the solar system,
where the planetary velocities decrease with distance from the Sun.

The revolutionary observational discovery of the 1970s and 1980s was that
galaxies do not behave this way.  \citet{rubin1978} showed optically, and
\citet{bosma1981} confirmed with 21\,cm measurements of neutral hydrogen (HI) extending to much
larger radii, that the RCs of spiral galaxies remain flat or even slightly
rising at radii well beyond the stellar disc, where the Keplerian prediction
would require a steep decline.  In Newtonian gravity this requires a
significant amount of mass distributed far beyond the optical boundary of
the galaxy---the dark matter halo.

Under spherical symmetry, equation~\eqref{eq:vc_spherical_mass} shows that a
flat RC $\vc = \mathrm{const}$ implies $M(<R) \propto R$, \ie\ a mass distribution with density
$\rho \propto R^{-2}$ extending to large radii (the singular isothermal
sphere, Section~\ref{sec:dm_profiles}).  More physically motivated profiles,
such as the NFW profile derived from $N$-body simulations
(Section~\ref{sec:nfw}), predict a slightly declining RC at large $R$ that is
nonetheless far shallower than Keplerian.

\subsection{The challenge of measuring the MW rotation curve}

Our position within the Galactic disc makes measuring the MW RC both uniquely
powerful and uniquely challenging.  On the one hand, we can access individual
stellar six-dimensional phase-space coordinates $(x,y,z,v_x,v_y,v_z)$---an
impossibility for any external galaxy.  On the other hand, three sources of
difficulty conspire against a straightforward measurement.

First, \textit{geometric degeneracy}: unlike an external galaxy viewed at a
known inclination, the MW RC cannot be read off directly from a velocity map.
The observed line-of-sight velocity of a star depends on its unknown distance,
its three-dimensional velocity, the solar motion, and the Galactic geometry.
Disentangling these factors requires either accurate distances (from parallax
or standard candles) or statistical methods that marginalise over the distance
distribution.

Second, \textit{asymmetric drift}: disc stars are not on perfectly circular
orbits; random motions cause the mean azimuthal velocity to lag behind the
local circular speed.  This asymmetric drift (AD) correction (Section~\ref{sec:asym_drift})
must be estimated from the velocity dispersion tensor and its gradients, and
is largest precisely where the RC measurement is most uncertain---the outer
disc at $R \gtrsim 15\kpc$.

Third, \textit{non-circular motions}: the Galactic bar, spiral arms,
tidal perturbations from the Sagittarius dwarf and the Large Magellanic Cloud
(LMC), and the warp of the outer disc all introduce non-axisymmetric velocity
components that can mimic or mask a genuine circular-speed variation.

Before \textit{Gaia}, these challenges meant that the MW virial mass was
known only to within a factor of several---$(0.5$--$3) \times 10^{12}\Msun$
\citep{bland-hawthorn2016}---despite decades of effort.  Pre-\textit{Gaia}
analyses of the inner MW rotation curve had nonetheless established robust
evidence for a non-baryonic dark component already at $R \lesssim R_0$
\citep{iocco2015}, complementing the kinematic and dynamical constraints
on the outer halo.

\subsection{The \textit{Gaia} revolution}

The \textit{Gaia} mission \citep{gaia2016}, launched in December 2013,
measures parallaxes and proper motions for $\sim 1.5$ billion stars, with
parallax precision $\sigma_\varpi \approx 0.01$--$0.1\,\mathrm{mas}$, and
radial velocities for $\sim 33$ million objects in DR3 \citep{katz2023}.  For
the first time, the full six-dimensional phase-space distribution of millions
of stars in the disc and halo is accessible, enabling RC determinations whose
statistical errors are smaller than the systematic uncertainties.  A broad
overview of the impact of Gaia on Galactic dynamics, including disequilibrium
effects, the bar, the phase spiral and the outer-halo response to the LMC,
is given in the review by \citet{vasiliev2025}.

The transformative power of Gaia DR3 is evident from the fact that it has
already produced several RC determinations extending to
$R > 20\kpc$, many finding a velocity decline---though disagreeing on its
magnitude and functional form.  The early \citet{eilers2019} analysis, using
$\sim 23\,000$ red giant branch (RGB) stars with APOGEE spectroscopy and
spectrophotometric distances,
found a gentle decline of slope $-1.7 \pm 0.1\kms\kpc^{-1}$ over
$5$--$25\kpc$ (with a reported systematic slope uncertainty of
$0.46\kms\kpc^{-1}$) and inferred a virial mass of
$\Mvir = (7.25 \pm 0.26) \times 10^{11}\Msun$ from an NFW fit, where the
quoted mass uncertainty is formal.  The
subsequent DR3-based analyses of \citet{wang2023}, \citet{jiao2023}, and
\citet{ou2024} all find a steeper decline at $R > 15\kpc$, with the extreme
interpretation of \citet{jiao2023} claiming a Keplerian decline and a total
mass nearly five times lower.

\subsection{Why this matters}

The stakes of the MW RC debate extend well beyond Galactic astronomy.  First,
the MW virial mass sets the normalisation for predictions of the satellite
galaxy population, the Local Group dynamics, and the merger history of our
Galaxy.  A factor-of-five reduction in the MW mass would imply far fewer
expected satellite galaxies, potentially significantly alleviating the
``missing satellites'' problem by bringing theoretical predictions into
closer agreement with the observed satellite population (though whether
the tension is fully resolved depends on feedback and reionisation
modelling at the low-mass end).  Second, the local DM
density $\rholocDM$ at the Sun's position, which for a halo of the form
$\rho(r) = \rho_s f(r/r_s)$ scales linearly with the density normalisation
$\rho_s$ at fixed geometry (it is the corresponding circular speed that
scales as $\sqrt{\rho_s}$), directly sets the signal expected in
direct-detection experiments such as LUX-ZEPLIN and XENON.  Third, the shape of the RC at
$R \approx 20$--$30\kpc$ constrains the DM halo profile, testing whether the
inner slope is cuspy (NFW) or cored and whether the halo is truncated at an
unexpectedly small radius.  Fourth, the RC decline is potentially in tension
with alternative gravity theories: the MW outer disc approaches the
low-acceleration MOND regime, where standard MOND predicts an asymptotically
\textit{flat} RC, making a Keplerian
decline particularly surprising.

\subsection{Structure of this review}

This article is organised as follows.  Section~\ref{sec:cbf} derives the
collisionless Boltzmann equation (CBE) from Liouville's theorem and reduces
it to the Jeans equations through a systematic velocity-moment expansion.
Section~\ref{sec:asym_drift} provides a detailed derivation of the asymmetric
drift formula and discusses the physical origin and observational estimation
of each contributing term.  Section~\ref{sec:dm_profiles} derives the
enclosed-mass expressions and circular-speed curves for the standard DM halo
profiles used in the literature.  Section~\ref{sec:baryons} describes the
baryonic mass models.  Section~\ref{sec:tracers} surveys the observational
tracers used to measure the MW RC and discusses the Gaia astrometric
precision.  Section~\ref{sec:gaia} presents the \textit{Gaia}-era RC
determinations in detail.  Section~\ref{sec:mass} discusses mass models and
the consequences of a Keplerian decline.  Section~\ref{sec:systematics}
critically examines systematic uncertainties.  Section~\ref{sec:altgrav}
treats alternative gravity theories, including a derivation of the MOND flat-RC
prediction.  Section~\ref{sec:indep} reviews independent dynamical constraints.
Section~\ref{sec:dm} discusses implications for DM density and particle
physics.  Section~\ref{sec:rar} examines the Radial Acceleration Relation.
Section~\ref{sec:discussion} discusses the three possible scenarios for
reconciling the observations, and Section~\ref{sec:conclusions} presents our
conclusions.  Four appendices collect useful unit conversions, the spherical
Jeans solution, the NFW potential, and the baryonic Tully-Fisher relation.

Throughout this review we adopt $R_0 \equiv \Rsun = 8.178\kpc$
\citep{gravity2019} for the Galactocentric distance of the Sun (the two
symbols are used interchangeably) and
$\vc(R_0) \approx 229$--$237\kms$ \citep{eilers2019,gravity2019,mroz2019}.
We use $G = 4.302 \times 10^{-3}\,\mathrm{pc}\,\Msun^{-1}(\kms)^2 =
4.302 \times 10^{-6}\,\mathrm{kpc}\,\Msun^{-1}(\kms)^2$.

\section{The Collisionless Boltzmann Equation}
\label{sec:cbf}

\subsection{The distribution function and Liouville's theorem}

The fundamental object describing the dynamical state of a stellar system is
the \textit{distribution function} (DF) $f(\bm{x}, \bm{v}, t)$, defined such
that $f\,\dif^3\bm{x}\,\dif^3\bm{v}$ gives the number of stars within the
phase-space volume element $\dif^3\bm{x}\,\dif^3\bm{v}$ at time $t$.  The
physical number density is recovered by integrating over velocities:
$n(\bm{x}, t) = \int f\,\dif^3\bm{v}$, and the total number of stars is
$N = \int f\,\dif^3\bm{x}\,\dif^3\bm{v}$.

For the smooth Galactic components, stars can be approximated as interacting
through the collective gravitational potential $\Phi(\bm{x}, t)$; direct
two-body encounters between individual
stars are negligible on timescales of interest because the two-body relaxation
time
\begin{equation}
  t_{\mathrm{relax}} \sim \frac{0.1\,N}{\ln N}\,t_{\mathrm{cross}}
  \label{eq:t_relax}
\end{equation}
far exceeds the Hubble time for the smooth Galactic disc and halo
($N \sim 10^{11}$,
$t_{\mathrm{cross}} \sim 300\,\mathrm{Myr}$, giving
$t_{\mathrm{relax}} \sim 10^{17}\,\mathrm{yr}$, some seven orders of
magnitude longer than the age of the Universe).  In this \textit{collisionless
approximation}, a fluid element in phase space moves along the trajectory of a
star, with $\dot{\bm{x}} = \bm{v}$ and $\dot{\bm{v}} = -\nabla\Phi$.

Liouville's theorem from classical mechanics states that the phase-space
density is conserved along any Hamiltonian trajectory.  For our stellar
system, this reads:
\begin{equation}
  \frac{\dif f}{\dif t}
  = \frac{\pp f}{\pp t}
  + \dot{\bm{x}} \cdot \frac{\pp f}{\pp \bm{x}}
  + \dot{\bm{v}} \cdot \frac{\pp f}{\pp \bm{v}}
  = 0.
  \label{eq:liouville}
\end{equation}
Substituting $\dot{\bm{x}} = \bm{v}$ and $\dot{\bm{v}} = -\nabla\Phi$
into equation~\eqref{eq:liouville} gives the
\textbf{collisionless Boltzmann equation} (CBE):
\begin{equation}
  \frac{\pp f}{\pp t}
  + \bm{v} \cdot \frac{\pp f}{\pp \bm{x}}
  - \nabla\Phi \cdot \frac{\pp f}{\pp \bm{v}}
  = 0.
  \label{eq:CBE}
\end{equation}
Equation~\eqref{eq:CBE} is the master equation for collisionless stellar
dynamics.  It is a first-order partial differential equation in
six-dimensional phase space, coupled through Poisson's equation
\begin{equation}
  \nabla^2\Phi = 4\pi G\,\rho_{\mathrm{tot}}
  \label{eq:poisson}
\end{equation}
to the \textit{total} gravitating mass density $\rho_{\mathrm{tot}}$.  Note
that the distribution function $f$ introduced above describes the stellar
population only, so $m_\star \int f\,\dif^3\bm{v}$ (where $m_\star$ is the
mean stellar mass) is the stellar mass density $\rho_\star$, which is just
one contribution to $\rho_{\mathrm{tot}}$; the latter must also include the
dark matter, the gas, and any additional mass components, each of which
obeys its own Boltzmann (or fluid) equation in the common potential $\Phi$.
In general, the CBE is
impossible to solve directly because it lives in a six-dimensional space.
Instead, one takes velocity moments to obtain the Jeans equations.

\subsection{Velocity moments and the Jeans equations}
\label{sec:jeans_derive}

We work in cylindrical coordinates $(R, \phi, z)$ with the convention
$\bm{v} = (v_R, v_\phi, v_z)$.  Define the tracer number density, mean
velocities, and velocity dispersion tensor as:
\begin{align}
  \nu(R, z) &\equiv \int f\,\dif^3\bm{v},
  \label{eq:nu_def}\\
  \langle v_i \rangle &\equiv \frac{1}{\nu}\int v_i f\,\dif^3\bm{v},
  \label{eq:mean_v}\\
  \sigma_{ij}^2 &\equiv
    \langle (v_i - \langle v_i\rangle)(v_j - \langle v_j\rangle)\rangle.
  \label{eq:sigma_tensor}
\end{align}
The diagonal elements $\sigma_{RR}^2$, $\sigma_{\phi\phi}^2$, and
$\sigma_{zz}^2$ are the variances of the radial, azimuthal, and vertical
velocity distributions; $\sigma_{Rz}^2$ is the covariance (cross-term)
between radial and vertical motions, which vanishes in the Galactic midplane
for a symmetric disc but is generally non-zero off the plane.

\subsubsection{Zeroth moment: the continuity equation}

Integrating the CBE~\eqref{eq:CBE} over all velocities and applying the
divergence theorem gives the continuity equation.  For steady state
($\pp/\pp t = 0$), axisymmetry ($\pp/\pp\phi = 0$), and no net radial or
vertical streaming ($\langle v_R\rangle = \langle v_z\rangle = 0$), the
equation is trivially satisfied and imposes no further constraint.

\subsubsection{First radial moment: derivation of the radial Jeans equation}

The more useful equation comes from multiplying the CBE~\eqref{eq:CBE} by
$v_R$ and integrating over all velocities.  In cylindrical coordinates, the
CBE in steady state reads:
\begin{equation}
  v_R\frac{\pp f}{\pp R}
  + \frac{v_\phi}{R}\frac{\pp f}{\pp\phi}
  + v_z\frac{\pp f}{\pp z}
  + \left(\frac{v_\phi^2}{R} - \frac{\pp\Phi}{\pp R}\right)\frac{\pp f}{\pp v_R}
  - \frac{v_R v_\phi}{R}\frac{\pp f}{\pp v_\phi}
  - \frac{\pp\Phi}{\pp z}\frac{\pp f}{\pp v_z}
  = 0.
  \label{eq:CBE_cylindrical}
\end{equation}
We multiply equation~\eqref{eq:CBE_cylindrical} by $v_R$ and integrate term
by term.  Using axisymmetry ($\pp/\pp\phi = 0$) and integration by parts on
the velocity derivatives (with surface terms vanishing), each term gives:

\medskip
\noindent\textit{Term 1:}
\begin{equation}
  \int v_R^2\,\frac{\pp f}{\pp R}\,\dif^3v
  = \frac{\pp}{\pp R}\int v_R^2 f\,\dif^3v
  = \frac{\pp(\nu\sigma_{RR}^2)}{\pp R},
  \label{eq:term1}
\end{equation}
where we used $\langle v_R\rangle = 0$ so $\int v_R^2 f\,\dif^3v = \nu\sigma_{RR}^2$.

\medskip
\noindent\textit{Term 2 ($\phi$-derivative):} vanishes by axisymmetry.

\medskip
\noindent\textit{Term 3:}
\begin{equation}
  \int v_R v_z\,\frac{\pp f}{\pp z}\,\dif^3v
  = \frac{\pp(\nu\sigma_{Rz}^2)}{\pp z}.
  \label{eq:term3}
\end{equation}

\medskip
\noindent\textit{Term 4 (centrifugal + gravitational):}
\begin{align}
  \int v_R\!\left(\frac{v_\phi^2}{R} - \frac{\pp\Phi}{\pp R}\right)
  \frac{\pp f}{\pp v_R}\,\dif^3v
  &= -\frac{\nu(\sigma_{\phi\phi}^2 + \langle v_\phi\rangle^2)}{R}
    + \nu\frac{\pp\Phi}{\pp R}.
  \label{eq:term4}
\end{align}

\medskip
\noindent\textit{Term 5:}
\begin{equation}
  -\int\frac{v_R^2 v_\phi}{R}\frac{\pp f}{\pp v_\phi}\,\dif^3v
  = \frac{\nu\sigma_{RR}^2}{R}.
  \label{eq:term5}
\end{equation}

Combining equations~\eqref{eq:term1}--\eqref{eq:term5}, using the
circular-speed relation $\pp\Phi/\pp R = \vc^2/R$ (the gravitational
acceleration is $-\pp\Phi/\pp R$, directed inward), and
$\langle v_\phi\rangle^2 + \sigma_{\phi\phi}^2 = \langle
v_\phi^2\rangle$, the \textbf{radial Jeans equation} emerges as:
\begin{equation}
  \vc^2(R)
  = \langle v_\phi\rangle^2
  + \sigma_{\phi\phi}^2 - \sigma_{RR}^2
  - \frac{R}{\nu}\frac{\pp(\nu\sigma_{RR}^2)}{\pp R}
  - \frac{R}{\nu}\frac{\pp(\nu\sigma_{Rz}^2)}{\pp z}.
  \label{eq:jeans_R}
\end{equation}
Equation~\eqref{eq:jeans_R} is the key relation used to infer the circular
speed $\vc(R)$ from observed stellar kinematics.  The circular speed is
recovered from the measured mean rotation $\langle v_\phi\rangle$ plus three
correction terms: the azimuthal--radial anisotropy term
$\sigma_{\phi\phi}^2 - \sigma_{RR}^2$, the radial pressure-gradient term
$-(R/\nu)\,\pp(\nu\sigma_{RR}^2)/\pp R$ (positive for density and dispersion
profiles that decline outward), and the vertical cross-term.  Each of these
terms requires measuring the velocity
dispersion tensor from the data, introducing systematic uncertainties that we
discuss extensively in Section~\ref{sec:systematics}.

\subsubsection{The vertical Jeans equation}

Multiplying the CBE~\eqref{eq:CBE_cylindrical} by $v_z$ and integrating in an
analogous fashion gives the \textbf{vertical Jeans equation}:
\begin{equation}
  \frac{\pp(\nu\sigma_{zz}^2)}{\pp z}
  + \frac{1}{R}\frac{\pp(R\nu\sigma_{Rz}^2)}{\pp R}
  = -\nu\frac{\pp\Phi}{\pp z}.
  \label{eq:jeans_z}
\end{equation}
Equation~\eqref{eq:jeans_z} relates the vertical velocity dispersion profile
to the vertical gravitational force, which near the midplane is dominated by
the local disc surface density.  It is the modern formulation of the classic
Oort limit problem: measuring $\sigma_{zz}(z)$ and $\nu(z)$ for a tracer
population determines $\rho_{\mathrm{tot}}(z)$ locally, providing an
independent determination of the local DM density complementary to the global
RC analysis.

\subsection{The spherical Jeans equation}

For a steady, non-rotating, approximately spherical tracer population in an
approximately spherical potential, the appropriate framework is the
\textbf{spherical Jeans equation}:
\begin{equation}
  \frac{\dif(\nu\sigma_r^2)}{\dif r}
  + \frac{2\beta_{\mathrm{ani}}\,\nu\sigma_r^2}{r}
  = -\nu\frac{G M(r)}{r^2},
  \label{eq:jeans_sph}
\end{equation}
where
$\beta_{\mathrm{ani}} = 1 -
(\sigma_\theta^2+\sigma_\phi^2)/(2\sigma_r^2)$ is the velocity anisotropy
parameter in spherical coordinates (reducing to
$1-\sigma_\theta^2/\sigma_r^2$ when the two tangential dispersions are
equal).  For constant
anisotropy $\beta_{\mathrm{ani}} = \beta_0$,
equation~\eqref{eq:jeans_sph} has the closed-form solution:
\begin{equation}
  \nu(r)\sigma_r^2(r)
  = r^{-2\beta_0}
    \int_r^\infty s^{2\beta_0}\,\nu(s)\,\frac{G M(s)}{s^2}\,\dif s,
  \label{eq:jeans_sph_sol}
\end{equation}
which is obtained by treating equation~\eqref{eq:jeans_sph} as a first-order
ODE and integrating with the boundary condition $\nu\sigma_r^2 \to 0$ as
$r \to \infty$.  Equation~\eqref{eq:jeans_sph_sol} is the tool used by
\citet{eadie2019} and \citet{posti2019} to infer the MW mass from globular
cluster kinematics.

The distinction between equations~\eqref{eq:jeans_R} and
\eqref{eq:jeans_sph} is not merely cosmetic.  \citet{klacka2025} noted that
the best-fit mass models of \citet{jiao2023} and \citet{ou2024} predict that
the spherically distributed DM halo accounts for $\gtrsim 80$--$90\%$ of the
radial gravitational acceleration at $R \gtrsim 20\kpc$, and argued that the
tracer population and the closure assumptions adopted in the cylindrical
analyses may then be inappropriate.  We stress that a spherical potential is
also axisymmetric, so the cylindrical Jeans
equations remain mathematically valid in such a potential; the substantive
question is whether the geometry, equilibrium, and phase-space structure
assumed for the \textit{tracer} are adequate.  We discuss this point in
Section~\ref{sec:jeans_inconsistency}.

\section{The Asymmetric Drift Correction}
\label{sec:asym_drift}

\subsection{Physical origin}

Stars in a galactic disc are not on perfectly circular orbits.  They are
perturbed repeatedly by molecular clouds, spiral arms, and the
non-axisymmetric bar potential, which exchange energy and angular momentum
with their orbits and raise their random motions.  This process, known as
disc heating,
increases the velocity dispersion components $\sigma_{RR}$,
$\sigma_{\phi\phi}$, and $\sigma_{zz}$ over time, while simultaneously
causing the mean azimuthal velocity $\langle v_\phi\rangle$ to lag below the
local circular speed $\vc$.

To see why this must be the case, consider a stellar population in Jeans
equilibrium in the radial direction.  The centripetal acceleration $\vc^2/R$
is balanced not only by the mean azimuthal motion $\langle v_\phi\rangle^2/R$
but also by the radial pressure gradient
$\sigma_{RR}^2\,\pp\ln(\nu\sigma_{RR}^2)/\pp R$.  In a disc where the density
and dispersion decrease outward, the pressure-gradient force is directed
outward and provides part of the radial support.  As a result,
$\langle v_\phi\rangle < \vc$: the population is in
\textit{sub-circular} rotation.  The lag
$\Delta v_\phi \equiv \vc - \langle v_\phi\rangle$ is the asymmetric drift
in its conventional definition.  Throughout this review we work instead
with the quadratic quantity $v_a$ defined through
$\vc^2 = \langle v_\phi\rangle^2 + v_a^2$, which arises naturally from the
radial Jeans equation; the two are related by
$v_a^2 = (\vc - \langle v_\phi\rangle)(\vc + \langle v_\phi\rangle)
\approx 2\vc\,\Delta v_\phi$, so that $v_a$ and $\Delta v_\phi$ are
\textit{not} numerically equal ($v_a \approx \sqrt{2\vc\,\Delta v_\phi}$
for $\Delta v_\phi \ll \vc$).  All formulae and numerical values in the
remainder of this review refer to the quadratic quantity $v_a$; where a
comparison with the conventional drift is useful we quote
$\Delta v_\phi \approx v_a^2/(2\vc)$ explicitly.

\subsection{Derivation of the asymmetric drift formula}

As anticipated above, we define $v_a$ through
\begin{equation}
  \vc^2 = \langle v_\phi\rangle^2 + v_a^2,
  \label{eq:vc_va}
\end{equation}
so that $v_a^2 = \vc^2 - \langle v_\phi\rangle^2 \approx 2\vc\,\Delta v_\phi$,
where $\Delta v_\phi = \vc - \langle v_\phi\rangle$ is the conventional
asymmetric drift.  Starting from the radial
Jeans equation~\eqref{eq:jeans_R}, we substitute
equation~\eqref{eq:vc_va} on the left-hand side and rewrite the
pressure-gradient term by separating
the logarithmic derivatives of $\nu$ and $\sigma_{RR}^2$:
\begin{align}
  \frac{1}{\nu}\frac{\pp(\nu\sigma_{RR}^2)}{\pp R}
  &= \sigma_{RR}^2\!\left(
    \frac{\pp\ln\nu}{\pp\ln R}\cdot\frac{1}{R}
    + \frac{\pp\ln\sigma_{RR}^2}{\pp\ln R}\cdot\frac{1}{R}\right).
  \label{eq:pressure_grad}
\end{align}
Substituting equation~\eqref{eq:pressure_grad} into \eqref{eq:jeans_R} and
solving for $v_a^2 = \vc^2 - \langle v_\phi\rangle^2$, one finds:
\begin{equation}
  v_a^2
  = -\sigma_{RR}^2\!\left(
    \frac{\pp\ln\nu}{\pp\ln R}
    + \frac{\pp\ln\sigma_{RR}^2}{\pp\ln R}
    + 1 - \kappa^2\right),
  \label{eq:va2_simple}
\end{equation}
where $\kappa^2 \equiv \sigma_{\phi\phi}^2 / \sigma_{RR}^2$ is the
axis ratio of the velocity ellipsoid.  Including the cross-term from
equation~\eqref{eq:jeans_R}:
\begin{equation}
  v_a^2
  = -\sigma_{RR}^2\!\left(
    \frac{\pp\ln\nu}{\pp\ln R}
    + \frac{\pp\ln\sigma_{RR}^2}{\pp\ln R}
    + 1 - \kappa^2\right)
  - \frac{R}{\nu}\frac{\pp(\nu\sigma_{Rz}^2)}{\pp z}.
  \label{eq:va2_full}
\end{equation}
Equations~\eqref{eq:va2_simple} and \eqref{eq:va2_full} are the practical
asymmetric drift formulae used in MW RC studies.  The observable
$\langle v_\phi\rangle$ is measured directly from the mean azimuthal velocity
of the tracer sample, and $\vc$ is then recovered from
equation~\eqref{eq:vc_va} as $\vc = \sqrt{\langle v_\phi\rangle^2 + v_a^2}$.

\subsection{Physical interpretation of each term}

Equation~\eqref{eq:va2_simple} contains three distinct contributions to the
asymmetric drift, each with a clear physical interpretation.

The first term, $-\sigma_{RR}^2\,\pp\ln\nu/\pp\ln R$, is the
\textit{density-gradient term}.  For a disc population with an exponential
density profile $\nu \propto e^{-R/h_R}$, one has
$\pp\ln\nu/\pp\ln R = -R/h_R < 0$, so this term contributes positively to
$v_a^2$.  In equilibrium there is no required net outward stellar flux.
Instead, stars observed away from their guiding centres sample eccentric
orbits; at larger radii they are preferentially near apocentre and rotate
more slowly than the local circular speed.  The declining tracer pressure
encodes this population-level support.

The second term, $-\sigma_{RR}^2\,\pp\ln\sigma_{RR}^2/\pp\ln R$, is the
\textit{dispersion-gradient term}.  If the velocity dispersion decreases
outward (the normal situation for a disc that has been heated more at early
times), this term is also positive: the outer disc is ``cooler'' than the
inner disc, so there is less pressure support per unit density in the outer
regions.  Through equation~\eqref{eq:va2_simple} this increases $v_a^2$ and
therefore lowers the mean azimuthal speed at fixed $\vc$.

\begin{figure}[htbp]
\centering
\includegraphics[width=\linewidth,keepaspectratio]{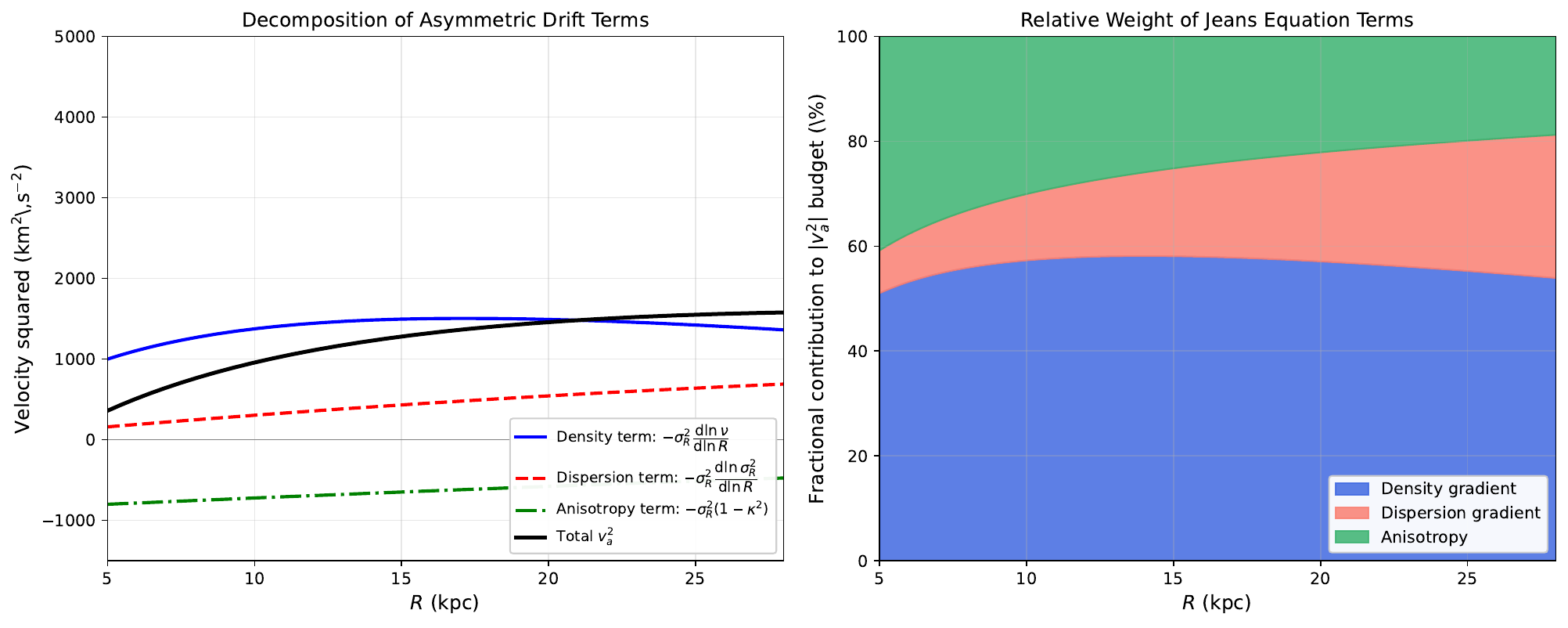}
\caption{\textbf{Left:} Decomposition of $v_a^2$
  (equation~\ref{eq:va2_simple}) into the density-gradient term (the largest
  positive contribution), the dispersion-gradient term (smaller positive
  contribution), and the anisotropy term ($-\sigma_{RR}^2(1-\kappa^2)$, which
  enters with the opposite sign for $\kappa^2 < 1$ and partially cancels
  the others), for a model RGB population matching the parameters of
  Section~\ref{sec:asym_drift}. The total $v_a^2$ (solid black) is the
  sum of the three terms and corresponds to $v_a \sim 20$--$40\kms$ across
  $R = 5$--$28\kpc$.
  \textbf{Right:} Fractional contributions to the $|v_a^2|$ budget. The
  density-gradient term provides $\sim 55\%$, the dispersion gradient
  $\sim 10$--$25\%$, and the anisotropy $\sim 20$--$40\%$, with the
  anisotropy fraction declining outward. None of the terms is negligible:
  the partial cancellation between the gradient terms and the
  (negative-sign) anisotropy term is what keeps $v_a$ in the
  illustrative $\sim 20$--$40\kms$ range, despite individual
  terms being significantly larger.}
\label{fig:jeans_terms}
\end{figure}

The third term, $-\sigma_{RR}^2(1 - \kappa^2)$, is the
\textit{anisotropy term}.  For a radially biased velocity ellipsoid
($\kappa < 1$, \ie\ $\sigma_{\phi\phi} < \sigma_{RR}$), one has
$1 - \kappa^2 > 0$ and this contribution to $v_a^2$ is therefore
\textit{negative}: it partially cancels the positive contributions of the
density-gradient and dispersion-gradient terms (cf.\ the decomposition in
Figure~\ref{fig:jeans_terms}).  Physically, it originates from the direct
appearance of $\sigma_{\phi\phi}^2 - \sigma_{RR}^2$ in the radial Jeans
equation~\eqref{eq:jeans_R}: a radially biased ellipsoid reduces the
azimuthal pressure relative to the radial one, lowering the drift required
for equilibrium at fixed gradients.  The term vanishes for an isotropic
distribution ($\kappa = 1$).

The fourth term (in equation~\ref{eq:va2_full}),
$-(R/\nu)\,\pp(\nu\sigma_{Rz}^2)/\pp z$, is the \textit{cross-term}.  It
arises from the tilt of the velocity ellipsoid relative to the cylindrical
coordinate axes, and its omission introduces a bias that grows with height
$|z|$ above the plane.

\subsection{Epicyclic approximation for $\kappa^2$}

In the \textit{epicyclic approximation}, valid when orbits are nearly circular,
the ratio of azimuthal to radial dispersion is determined by the ratio of the
epicyclic frequency $\kappa_{\mathrm{epi}}$ to the angular frequency
$\Omega = \vc/R$:
\begin{equation}
  \kappa^2 \equiv \frac{\sigma_{\phi\phi}^2}{\sigma_{RR}^2}
  = \frac{\kappa_{\mathrm{epi}}^2}{4\Omega^2},
  \label{eq:kappa2_epi}
\end{equation}
where the epicyclic frequency is:
\begin{equation}
  \kappa_{\mathrm{epi}}^2
  = R\frac{\dif\Omega^2}{\dif R} + 4\Omega^2
  = \frac{1}{R}\frac{\dif\vc^2}{\dif R} + \frac{2\vc^2}{R^2}.
  \label{eq:kappa_epi}
\end{equation}
For a flat RC ($\dif\vc/\dif R = 0$), $\kappa_{\mathrm{epi}} = \sqrt{2}\,\Omega$
and $\kappa^2 = 1/2$.  For a Keplerian RC ($\vc \propto R^{-1/2}$, so
$\dif\vc^2/\dif R = -\vc^2/R$):
\begin{equation}
  \kappa_{\mathrm{epi}}^2
  = \frac{1}{R}\!\left(-\frac{\vc^2}{R}\right) + \frac{2\vc^2}{R^2}
  = \frac{\vc^2}{R^2}
  = \Omega^2,
\end{equation}
giving $\kappa^2 = 1/4$.  Holding the tracer-density and dispersion gradients
fixed, the term $1-\kappa^2$ is then larger, so the bracket in
equation~\eqref{eq:va2_simple} is less negative and $v_a^2$ is
\textit{smaller} than for the flat-RC value $\kappa^2=1/2$.  This creates a subtle
circularity: if one assumes the flat-RC value $\kappa^2 = 1/2$ when computing
the correction in a region where the true potential is closer to Keplerian,
the correction is overestimated and $\vc$ is consequently biased too high.

\subsection{Magnitude of the asymmetric drift in the outer disc}

For an RGB tracer in the outer MW disc, \citet{eilers2019} and
\citet{ou2024} measure $\sigma_{RR} \sim 30$--$40\kms$ at
$R \sim 20$--$25\kpc$. The density-gradient term
$\pp\ln\nu/\pp\ln R \approx -2$ to $-3$ in this radial range (corresponding
to an illustrative effective radial scale length
$h_R \sim 7$--$10\kpc$), the dispersion gradient
$\pp\ln\sigma_{RR}^2/\pp\ln R \approx -(0.3\text{--}0.5)$, and the
anisotropy is $\kappa^2 \approx 0.5$ (epicyclic value for an
approximately flat RC).

Substituting into equation~\eqref{eq:va2_simple}, we find:
\begin{align}
  v_a^2
  &\approx -\sigma_{RR}^2\bigl[(-2.5) + (-0.4) + (1 - 0.5)\bigr]
   = \sigma_{RR}^2 \times 2.4,
\end{align}
which for $\sigma_{RR} \approx 35\kms$ yields a quadratic asymmetric drift of 
$v_a \approx 54\kms$. The corresponding conventional linear drift lag is 
$\Delta v_\phi \approx v_a^2/(2\vc) \approx 6.4\kms$ (assuming $\vc \approx 230\kms$). 
This toy calculation is illustrative rather than an observational bound:
the individual logarithmic gradients of the tracer density and velocity
dispersion, and the neglected cross-term, are poorly constrained in the
outer disc.  Nonetheless, the sign structure is robust: because
both $\pp\ln\nu/\pp\ln R < 0$ and $\pp\ln\sigma_{RR}^2/\pp\ln R < 0$, the 
density and dispersion gradients enter with the same positive sign, while the 
anisotropy term $(1-\kappa^2) > 0$ provides the opposite contribution, reducing 
the net bracket value (cf.\ Figure~\ref{fig:jeans_terms}).

For plausible outer-disc tracer profiles the quadratic correction is
$v_a\sim30$--$55\kms$ (the toy value above lies at the upper end),
it propagates into the inferred circular speed weakly, since
$\vc^2 = \langle v_\phi\rangle^2 + v_a^2$ implies
$\delta \vc \approx (v_a / \vc)\,\delta v_a$. For this range of $v_a$
and $\vc \approx 230\kms$, even a $30\%$ error in $v_a$ shifts $\vc$ by
only $\sim 1$--$4\kms$, below the $\sim 30\kms$ outer-disc decline
claimed by \citet{jiao2023}. The asymmetric drift correction \textit{by
itself} is therefore not a dominant systematic in the outer-disc RC
inference. It nonetheless deserves attention because the same tracer
density and dispersion gradients that enter $v_a$ also appear in the
radial pressure-support term of the Jeans equation
(equation~\ref{eq:jeans_R}); a coordinated mis-estimation of the tracer
profile can therefore bias both $v_a$ \textit{and} the directly-inferred
$\vc^2$ in a correlated way that is harder to constrain from kinematics
alone. This coordinated bias, together with the distance-calibration and
non-equilibrium effects discussed in Section~\ref{sec:systematics}, is
where most of the systematic budget actually resides.

\section{Dark Matter Halo Profiles}
\label{sec:dm_profiles}

\subsection{The NFW profile}
\label{sec:nfw}

The Navarro--Frenk--White (NFW) profile \citep{navarro1997} is the canonical
DM halo model emerging from collisionless $N$-body simulations of structure
formation in $\Lambda$CDM:
\begin{equation}
  \rho_{\mathrm{NFW}}(r)
  = \frac{\rho_s}{(r/r_s)(1 + r/r_s)^2}.
  \label{eq:nfw_rho}
\end{equation}
At small radii $r \ll r_s$, the density diverges as $\rho \propto r^{-1}$
(a cusp); at large radii $r \gg r_s$, it falls as $\rho \propto r^{-3}$.

\subsubsection{Enclosed mass and its derivation}

We compute $M(r) = \int_0^r 4\pi r'^2 \rho(r') \dif r'$:
\begin{align}
  M_{\mathrm{NFW}}(r)
  &= 4\pi\rho_s r_s^3
    \int_0^{r/r_s}\frac{x\,\dif x}{(1+x)^2}.
  \label{eq:nfw_mass_integral}
\end{align}
The integral in equation~\eqref{eq:nfw_mass_integral} is evaluated by partial
fractions:
\begin{align}
  \int_0^s\frac{x\,\dif x}{(1+x)^2}
  &= \int_0^s\!\left[\frac{1}{1+x} - \frac{1}{(1+x)^2}\right]\dif x
   = \ln(1+s) - \frac{s}{1+s},
  \label{eq:nfw_integral}
\end{align}
where we used $[(1+x)^{-1}]_0^s = (1+s)^{-1} - 1$.  Therefore:
\begin{equation}
  M_{\mathrm{NFW}}(r)
  = 4\pi\rho_s r_s^3\!\left[\ln\!\left(1+\frac{r}{r_s}\right)
    - \frac{r/r_s}{1+r/r_s}\right].
  \label{eq:nfw_mass}
\end{equation}

\subsubsection{NFW circular speed}

Using equations~\eqref{eq:vc_basic} and \eqref{eq:nfw_mass}:
\begin{equation}
  v_{c,\mathrm{NFW}}^2(r)
  = \frac{4\pi G\rho_s r_s^3}{r}
    \left[\ln\!\left(1+\frac{r}{r_s}\right) - \frac{r/r_s}{1+r/r_s}\right].
  \label{eq:nfw_vc}
\end{equation}
In terms of the dimensionless radius $x = r/R_{200}$ and the virial circular
speed $V_{200}^2 = GM_{200}/R_{200}$:
\begin{equation}
  \frac{v_{c,\mathrm{NFW}}^2(x)}{V_{200}^2}
  = \frac{1}{x}\,
    \frac{\ln(1+cx) - cx/(1+cx)}{\ln(1+c) - c/(1+c)},
  \label{eq:nfw_vc_dimless}
\end{equation}
where $c = R_{200}/r_s$ is the concentration parameter (equivalently, with
$x' = r/r_s$, $v_c^2(x')/V_{200}^2 = (c/x')\,[\ln(1+x') -
x'/(1+x')]/[\ln(1+c) - c/(1+c)]$).  The peak of
$v_{c,\mathrm{NFW}}$ occurs at $r_{\mathrm{max}} \approx 2.16\,r_s$.

\subsubsection{Gravitational potential}

Integrating $\dif\Phi/\dif r = GM_{\mathrm{NFW}}(r)/r^2$ with
$\Phi(\infty)=0$:
\begin{equation}
  \Phi_{\mathrm{NFW}}(r)
  = -\frac{4\pi G\rho_s r_s^3}{r}
    \ln\!\left(1+\frac{r}{r_s}\right).
  \label{eq:nfw_phi}
\end{equation}

For an illustrative MW NFW model with $r_s\simeq14.8\kpc$,
$R_0 = 8.178\kpc \approx 0.55\,r_s$.  At
$R \approx 20\kpc \approx 1.35\,r_s$ the Galaxy is still inside
the radius $r_{\mathrm{max}} \approx 2.16\,r_s \approx 32\kpc$ at which the
halo-only $v_{c,\mathrm{NFW}}$ curve peaks---the total RC (baryons + halo)
has already begun declining, but at a rate much slower than Keplerian.  A genuine Keplerian decline at these radii would
require a strongly truncated or unusually light and concentrated halo; the
numerical scale radius and $M_{200}$ inferred from a finite radial interval
remain sensitive to the baryonic model and the concentration prior.

\subsubsection{Virial mass and concentration}

The virial radius $R_{200}$ is defined as the radius within which the mean
enclosed density equals $200\rho_{\mathrm{crit}}$:
\begin{equation}
  M_{200} = \frac{800\pi}{3}\rho_{\mathrm{crit}}R_{200}^3.
  \label{eq:M200_def}
\end{equation}
The characteristic density is then:
\begin{equation}
  \rho_s = \frac{200}{3}\,\frac{c^3}{\ln(1+c) - c/(1+c)}\,\rho_{\mathrm{crit}}.
  \label{eq:rhos_c}
\end{equation}
For a Planck-2018 cosmology ($h = 0.674$, $\Omega_m = 0.315$):
$\rho_{\mathrm{crit}} = 2.775 \times 10^{11}\,h^2\,\Msun\,\mathrm{Mpc}^{-3}
\simeq 1.26 \times 10^{11}\Msun\,\mathrm{Mpc}^{-3}$.
MW-mass halos have $c \approx 10$--$15$ \citep{dutton2014}, corresponding to
$r_s \approx 14$--$21\kpc$ for $M_{200} \approx 10^{12}\Msun$.

\begin{figure}[htbp]
\centering
\includegraphics[width=\linewidth,keepaspectratio]{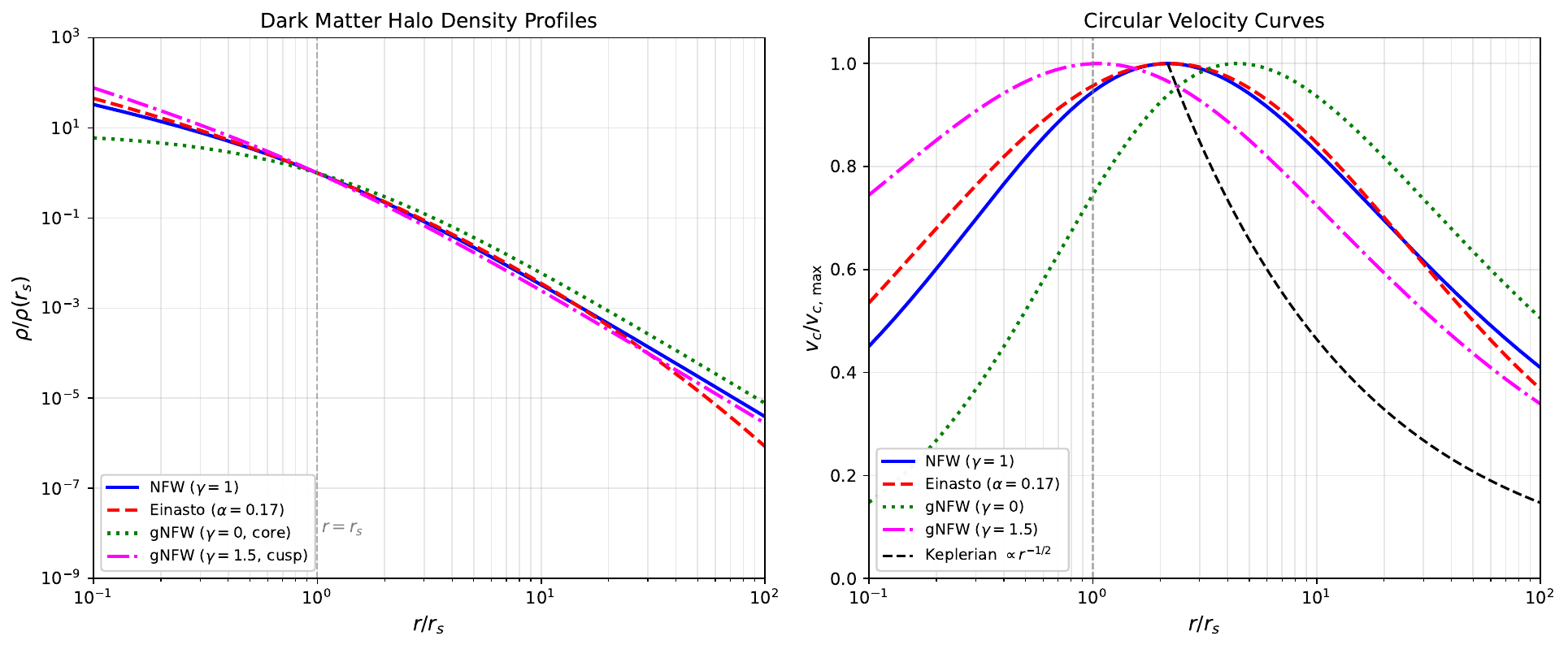}
\caption{\textbf{Left panel:} Normalised density profiles for four DM halo
  models versus $r/r_s$.  The NFW profile (blue solid, $\gamma=1$) has an
  inner cusp $\rho \propto r^{-1}$; the cored gNFW (green dotted, $\gamma=0$)
  has a flat central density; the steep gNFW (magenta dot-dash, $\gamma=1.5$)
  diverges faster; and the Einasto profile (red dashed, $\alpha=0.17$) falls
  more steeply than NFW at large radii.
  \textbf{Right panel:} Circular velocity curves for the same profiles,
  normalised to their peak values.  The Keplerian $\propto r^{-1/2}$ decline
  (black dashed) is shown for comparison.  None of the standard DM profiles
  reaches a Keplerian decline at $r \sim r_s$.}
\label{fig:dm_profiles}
\end{figure}

Figure~\ref{fig:dm_profiles} compares the shapes and circular-speed curves
of the standard spherical halo profiles discussed in this section.

\subsection{The Einasto profile}
\label{sec:einasto}

The Einasto profile \citep{einasto1965} provides a marginally better fit to
high-resolution $N$-body simulations than NFW \citep{navarro2010}:
\begin{equation}
  \rho_{\mathrm{Ein}}(r)
  = \rho_{-2}\exp\!\left\{-\frac{2}{\alpha}
    \left[\left(\frac{r}{r_{-2}}\right)^\alpha - 1\right]\right\}.
  \label{eq:einasto_rho}
\end{equation}
The enclosed mass requires the lower incomplete gamma function
$\gamma(a,x) = \int_0^x t^{a-1} e^{-t}\,\dif t$:
\begin{equation}
  M_{\mathrm{Ein}}(r)
  = 4\pi\rho_{-2} r_{-2}^3\,e^{2/\alpha}
    \left(\frac{\alpha}{2}\right)^{3/\alpha}\!\frac{1}{\alpha}\,
    \gamma\!\left(\frac{3}{\alpha},
    \frac{2}{\alpha}\left(\frac{r}{r_{-2}}\right)^\alpha\right).
  \label{eq:einasto_mass}
\end{equation}
For $\alpha \approx 0.17$--$0.20$ the Einasto profile is similar to NFW
at intermediate radii but falls faster than any power law at large $r$ and
has a finite total mass.  This commonly simulated range should not be
confused with the fit by \citet{ou2024}: their declining RC gave
$\alpha=0.91^{+0.04}_{-0.05}$ and, in their parametrisation
$\rho\propto\exp[-(r/r_s)^\alpha]$, a scale radius
$r_s=3.86^{+0.35}_{-0.38}\kpc$ (corresponding to
$r_{-2}=(2/\alpha)^{1/\alpha}r_s\approx9\kpc$ in the notation of
equation~\ref{eq:einasto_rho}), with
$M_{200}=1.81^{+0.06}_{-0.05}\times10^{11}\Msun$ under their adopted
baryonic and halo models.

\subsection{Generalised NFW profile}
\label{sec:gnfw}

The generalised NFW (gNFW) profile allows the inner density slope to vary as
a free parameter:
\begin{equation}
  \rho_{\mathrm{gNFW}}(r)
  = \frac{\rho_s}{(r/r_s)^\gamma(1 + r/r_s)^{3-\gamma}},
  \quad 0 \le \gamma \le 2,
  \label{eq:gnfw_rho}
\end{equation}
with $\gamma = 1$ recovering the standard NFW profile~\eqref{eq:nfw_rho}.
The enclosed mass is:
\begin{align}
  M_{\mathrm{gNFW}}(r)
  &= 4\pi\rho_s r_s^3
    \int_0^{r/r_s}\frac{x^{2-\gamma}}{(1+x)^{3-\gamma}}\,\dif x.
  \label{eq:gnfw_mass_integral}
\end{align}
For the cored case $\gamma = 0$:
\begin{align}
  \int_0^s\frac{x^2}{(1+x)^3}\,\dif x
  &= \int_0^s\!\left[\frac{1}{1+x} - \frac{2}{(1+x)^2}
    + \frac{1}{(1+x)^3}\right]\dif x\nonumber\\
  &= \ln(1+s)
    + \frac{2}{1+s} - \frac{1}{2(1+s)^2} - \frac{3}{2}.
  \label{eq:gnfw_cored_integral}
\end{align}
Hydrodynamic simulations including star formation and supernova feedback can
convert cuspy ($\gamma = 1$) into cored ($\gamma = 0$) profiles
\citep{pontzen2014}.  In the notation of equation~\eqref{eq:gnfw_rho},
\citet{ou2024} obtained a nearly cored fit,
$\gamma=0.026^{+0.042}_{-0.019}$ (i.e.\ pinned against the lower edge of
the prior) and $r_s=5.26^{+0.15}_{-0.11}\kpc$, with
$M_{200}=6.94^{+0.12}_{-0.11}\times10^{11}\Msun$.  Its reduced
$\chi^2=7.79$ was substantially worse than the Einasto fit
($\chi^2_\nu=2.97$), so the authors did not recommend the gNFW result as
their preferred mass model.

\subsection{Burkert profile}
\label{sec:burkert}

The Burkert profile \citep{burkert1995} was introduced to fit dwarf galaxy
RCs, and was subsequently shown to describe the halos of spirals of all
luminosities, with core radius and central density obeying tight scaling
relations \citep{salucci2000}; for a review of the resulting picture of
cored DM distributions in galaxies see \citet{salucci2019}.  Its
application to the MW, with a comparison against NFW on a compiled RC,
is given by \citet{nesti2013}:
\begin{equation}
  \rho_{\mathrm{Bk}}(r)
  = \frac{\rho_0 r_0^3}{(r + r_0)(r^2 + r_0^2)},
  \label{eq:burkert_rho}
\end{equation}
with enclosed mass:
\begin{equation}
  M_{\mathrm{Bk}}(r)
  = \pi\rho_0 r_0^3\!\left[
    \ln\!\left(1 + \frac{r^2}{r_0^2}\right)
    + 2\ln\!\left(1 + \frac{r}{r_0}\right)
    - 2\arctan\!\left(\frac{r}{r_0}\right)\right].
  \label{eq:burkert_mass}
\end{equation}
At large radii, $M_{\mathrm{Bk}} \propto \ln(r/r_0)$ and the RC declines
sub-Keplerian, similar to the NFW asymptotic behaviour.

\subsection{Singular isothermal sphere}
\label{sec:sis}

For completeness, the singular isothermal sphere (SIS) with
$\rho(r) = \sigma_v^2/(2\pi G r^2)$ gives $M(r) = (2\sigma_v^2/G)r$ and
$\vc = \sqrt{2}\,\sigma_v = \mathrm{const}$---a perfectly flat RC.  The SIS
is unphysical (infinite total mass) but is widely used as the simplest model
encoding a flat RC.

\section{Baryonic Mass Models}
\label{sec:baryons}

The total circular speed is decomposed into baryonic and DM contributions:
\begin{equation}
  \vc^2(R)
  = v_{c,\mathrm{bary}}^2(R) + v_{c,\mathrm{DM}}^2(R),
  \label{eq:vc_decomp}
\end{equation}
with
$v_{c,\mathrm{bary}}^2 = v_{c,\mathrm{bulge}}^2 + v_{c,\mathrm{thin}}^2
+ v_{c,\mathrm{thick}}^2 + v_{c,\mathrm{gas}}^2$.

\subsection{Hernquist bulge}

The \citet{hernquist1990} profile for the MW bulge/bar:
\begin{equation}
  \rho_{\mathrm{Hern}}(r)
  = \frac{M_b}{2\pi}\frac{a}{r(r+a)^3},
  \quad
  M_{\mathrm{Hern}}(r) = M_b\frac{r^2}{(r+a)^2},
  \label{eq:hernquist}
\end{equation}
giving:
\begin{equation}
  v_{c,\mathrm{bulge}}^2(R)
  = \frac{G M_b R}{(R + a)^2}.
  \label{eq:vc_bulge}
\end{equation}
Typical parameters: $M_b \approx (0.9$--$1.5) \times 10^{10}\Msun$,
$a \approx 0.6$--$0.8\kpc$ \citep{pouliasis2017,desalas2019}.

\subsection{Miyamoto--Nagai disc}

The \citet{miyamoto1975} potential:
\begin{equation}
  \Phi_{\mathrm{MN}}(R, z)
  = -\frac{GM_d}{\sqrt{R^2 + \bigl(a_d + \sqrt{z^2 + b_d^2}\bigr)^2}},
  \label{eq:mn_phi}
\end{equation}
with circular speed in the equatorial plane:
\begin{align}
  v_{c,\mathrm{MN}}^2(R)
  &= R\!\left.\frac{\pp\Phi_{\mathrm{MN}}}{\pp R}\right|_{z=0}
   = \frac{GM_d\,R^2}{\bigl[R^2 + (a_d + b_d)^2\bigr]^{3/2}}.
  \label{eq:vc_mn}
\end{align}
Thin disc: $M_{\mathrm{thin}} \approx 3$--$4 \times 10^{10}\Msun$,
$a_{\mathrm{thin}} \approx 3\kpc$, $b_{\mathrm{thin}} \approx 0.28\kpc$.
Thick disc: $M_{\mathrm{thick}} \approx 0.5$--$1 \times 10^{10}\Msun$,
$a_{\mathrm{thick}} \approx 3.5\kpc$, $b_{\mathrm{thick}} \approx 0.9\kpc$.

\subsection{Exponential disc}

For a razor-thin exponential disc with surface density
$\Sigma(R) = \Sigma_0 e^{-R/h_R}$, the exact circular speed is
\citep{freeman1970}:
\begin{equation}
  v_{c,\mathrm{exp}}^2(R)
  = 4\pi G\Sigma_0 h_R\,y^2
    \bigl[I_0(y)K_0(y) - I_1(y)K_1(y)\bigr],
  \quad y = \frac{R}{2h_R},
  \label{eq:vc_expdisk}
\end{equation}
where $I_n$ and $K_n$ are modified Bessel functions.

\subsection{Total baryonic budget}

The total baryonic mass of the MW is
$M_{\rm bary} \approx (0.6$--$1.0) \times 10^{11}\Msun$
\citep{bland-hawthorn2016}.  At $R = R_0$, the baryonic contribution amounts
to $v_{c,\rm bary}(R_0) \approx 170$--$190\kms$, leaving a DM contribution
of $v_{c,\rm DM}(R_0) \approx 130$--$150\kms$; since the contributions
add in quadrature, DM supplies $v_{c,\rm DM}^2/\vc^2\sim30$--$40\%$ of the
radial force at the solar radius.
In standard MW mass models the DM halo generally dominates the radial force
by $R\sim15$--$20\kpc$, although the precise transition depends on the
baryonic model.

\section{Observational Tracers}
\label{sec:tracers}

\subsection{The tangent-point method for $R < R_0$}

For Galactic longitudes $|l| < 90^\circ$, the maximum line-of-sight velocity
at longitude $l$ occurs at the subcentral (tangent) point
$R_{\rm tan} = \Rsun|\sin l|$:
\begin{equation}
  \vc(R_{\rm tan}) = v_{\rm LSR,max}(l) + \vc(R_0)|\sin l|,
  \label{eq:tangent}
\end{equation}
where $v_{\rm LSR,max}(l)$ is the maximum line-of-sight velocity observed
at that longitude, measured with respect to the Local Standard of Rest
(LSR), i.e.\ corrected for the Sun's peculiar motion.
This method is restricted to $R < R_0$ and is affected by non-circular
motions near the bar ($|l| \lesssim 30^\circ$).

\subsection{Stellar populations used in \textit{Gaia}-era studies}

Schematically, the transformation from heliocentric observables to
Galactocentric cylindrical velocities is
\begin{equation}
  \begin{pmatrix}v_R\\ v_\phi\\ v_z\end{pmatrix}
  = \mathbf{T}(l, b, d)
  \begin{pmatrix}v_{\rm los}\\
  4.74047\,\mu_l^\ast d\\ 4.74047\,\mu_b d\end{pmatrix}
  +\bm{v}_{\odot\rightarrow{\rm GC}},
  \label{eq:velocity_transform}
\end{equation}
where $v_{\rm los}$ is the heliocentric line-of-sight velocity,
$(\mu_l^\ast,\mu_b)$ are the proper motions in Galactic coordinates in
mas\,yr$^{-1}$ (with $\mu_l^\ast=\mu_l\cos b$), $d$ is the distance in kpc,
velocities are in km\,s$^{-1}$, the factor $4.74047\kms$ converts
$\mathrm{mas\,yr^{-1}\times kpc}$ into km\,s$^{-1}$, $\mathbf{T}$ is the
rotation matrix that depends on sky position, and
$\bm{v}_{\odot\rightarrow{\rm GC}}$ accounts for the Solar motion and the
Local Standard of Rest.

\textit{Luminous RGB stars near the tip} reach absolute magnitudes
$M_K \approx -6$ to $-7$,
detectable to $d \sim 30\kpc$.  They require a large asymmetric drift
correction because, as intermediate-to-old-age stars, they have experienced
significant disc heating.

\textit{Classical Cepheids} follow the Leavitt period--luminosity relation.
Their individual distances are commonly precise at the few-percent to
$\sim10\%$ level, depending on calibration and extinction treatment; the
Gaia DR3 analysis of \citet{feng2026}, for example, reports a typical
uncertainty of about $7\%$.  As young, kinematically cold stars they require
only a small asymmetric-drift correction (about $1.6\kms$ in that analysis),
making them valuable complements to dynamically hotter RGB tracers.

\textit{VLBI masers} in star-forming regions provide geometric parallax
accuracies $\sigma_d/d < 3\%$.  The BeSSeL \citep{reid2014} and VERA
\citep{honma2012} surveys have used $\sim 100$ masers to constrain
$\Rsun = 8.34 \pm 0.16\kpc$ and $\vc(R_0) = 240 \pm 8\kms$.

\subsection{Gaia astrometric precision and systematics}

Gaia DR3 parallax precision depends on magnitude, colour, position, scanning
law, and source quality; a single polynomial in $G$ does not describe the
catalogue.  Representative uncertainties range from a few
$10\,\mu\mathrm{as}$ for bright, well-behaved sources to several
$10^{-1}\,\mathrm{mas}$ near the faint limit \citep{katz2023}.
A non-trivial parallax zero-point offset of
$\Delta\varpi \approx -17\,\mu\mathrm{as}$ was identified in Gaia DR3
\citep{lindegren2021}.  After the \citet{lindegren2021} correction, residual
spatially and colour-dependent systematics at the several-$\mu$as level can
remain.  A $5\,\mu\mathrm{as}$ residual is $5\%$ of the parallax of a source
at $10\kpc$, but its effect on an RC estimate depends on the selection
function, distance inference, viewing geometry, and velocity model; it
cannot in general be converted into a universal $5\%$ velocity bias.

\section{The \textit{Gaia} Era: RC Determinations and the Decline}
\label{sec:gaia}

\subsection{Pre-DR3 baseline: Eilers et al.\ (2019)}

\citet{eilers2019} established the pre-DR3 benchmark using $23\,000$ RGB
stars with APOGEE spectroscopy.  Their main results:
\begin{align*}
  \vc(R_0) &= 229.0 \pm 0.2\kms\;(\text{formal}),\\
  \dif\vc/\dif R &= -1.7 \pm 0.1\;(\text{formal})
    \pm0.46\;(\text{syst.})\kms\kpc^{-1},\\
  \Mvir &= (7.25 \pm 0.26) \times 10^{11}\Msun,\quad c = 12.8 \pm 0.3.
\end{align*}
These results apply over $R=5$--$25\kpc$.  The NFW virial mass was derived by
fitting equation~\eqref{eq:nfw_vc_dimless} to the RC; its quoted uncertainty
is formal, while the authors estimated a $2$--$5\%$ systematic uncertainty
in the circular-speed normalisation.

\subsection{Wang et al.\ (2023): extending to 28\,kpc}

\citet{wang2023} applied the Lucy iterative deconvolution method to the full
Gaia DR3 source catalogue, statistically correcting the observed proper-motion
distribution for parallax errors.  Their tabulated RC (anticentre region,
$|z|<3\kpc$, $\Rsun=8.34\kpc$) declines from $\vc \approx 221$--$224\kms$
at $R=9.5$--$13.5\kpc$ to $\vc \approx 175\kms$ at $27.5\kpc$, with a
fitted slope of $-(2.3\pm0.2)\kms\kpc^{-1}$.  An important
finding is a significant north-south kinematic asymmetry at
$R \sim 20$--$25\kpc$---stars above and below the plane show mean radial
velocities differing by $\sim 10$--$20\kms$---directly indicating
non-axisymmetric or non-equilibrium kinematics, violating one of the key
assumptions of the Jeans equation~\eqref{eq:jeans_R}.  A further caveat
specific to this determination is that the Lucy deconvolution yields
\textit{statistical} distances for the ensemble rather than individual
stellar distances: the inversion is sensitive to the adopted selection
function and to the well-known biases of Lucy-type iterations (noise
amplification and dependence on the stopping criterion), and its distance
scale cannot be validated star by star against high-precision independent
calibrators such as Cepheids or globular clusters.  The \citeauthor{wang2023}
RC should therefore be regarded as somewhat less directly anchored in
distance than the spectrophotometric RGB-based determinations.

\subsection{Jiao et al.\ (2023): detection of a Keplerian decline}
\label{sec:jiao2023}

\citet{jiao2023} re-derived the RC of \citet{wang2023} with a revised
binning and a full systematic error budget, and compared it with a
preliminary version of \citet{ou2024} (circulated as Ou et al.\ 2023 on
arXiv at the time of submission), after carefully reconciling their
different distance scales; the RC of \citet{zhou2023} was set aside on
the grounds that its distances appeared systematically larger than those
of other spectrophotometric scales, and \citet{eilers2019}
serves as the \textit{Gaia} DR2 baseline rather than as an input to the
combination.  Both RCs decline with a very similar linear slope,
$\beta = -(2.18 \pm 0.23)\kms\kpc^{-1}$ for their own measurement and
$\beta = -(2.22 \pm 0.20)\kms\kpc^{-1}$ for that of \citet{ou2024}.
They then fit a power-law model $\vc \propto R^\gamma$ to the data at
$R > 19\kpc$, finding $\gamma = -0.47 \pm 0.15$ for their RC and
$\gamma = -0.56^{+0.23}_{-0.22}$ for the \citet{ou2024} RC, both
consistent with Keplerian ($\gamma = -0.5$) at $1\sigma$.  A flat RC
($\gamma = 0$) is excluded at the $3\sigma$ level.  It is worth
stressing that the Keplerian behaviour is not a feature of the
\citeauthor{jiao2023} reanalysis alone: although neither
\citet{wang2023} nor \citet{ou2024} explicitly claim a Keplerian
decline, the outer slopes of their own RCs, taken at face value, are
consistent with one, and the power-law fit to the \citeauthor{ou2024}
data quoted above makes this explicit.

Regarding the exclusion of \citet{zhou2023}, we note that the argument
of \citeauthor{jiao2023} (their Fig.~B1) rests on a comparison with the
distance scales of Wang et al.\ (2016, as updated with Gaia EDR3 priors),
Hogg et al.\ (2019) and StarHorse (Queiroz et al.\ 2023), all of which are
somewhat shorter than the \citeauthor{zhou2023} scale for $R>10\kpc$.  Since all of these are spectrophotometric
distances, the offset is more naturally read as a difference among
calibrations than as proof that one of them is biased; the
\citeauthor{zhou2023} scale was itself calibrated against, and agrees well
with, the independent distances of globular clusters (their Figs.~3
and~8).  We therefore describe the situation as a genuine, unresolved
spread in distance calibrations---which propagates linearly into $\vc$
(Section~\ref{sec:dist_syst})---rather than as an established
overestimate of the \citeauthor{zhou2023} distances.

Under the spherical interpretation, the corresponding dynamical mass from
equation~\eqref{eq:vc_spherical_mass}
evaluated at $R_{\rm kep} = 19\kpc$ with $\vc \approx 215\kms$:
\begin{align}
  M_{\rm tot}
  &= \frac{\vc^2(R_{\rm kep})\,R_{\rm kep}}{G}
   = \frac{(215\kms)^2 \times 19\kpc}
     {4.302 \times 10^{-6}\,\mathrm{kpc}\,\Msun^{-1}(\kms)^2}
  \approx 2.04 \times 10^{11}\Msun.
  \label{eq:jiao_mass}
\end{align}
This spherical-equivalent value is consistent with
$M_{\rm bary} \approx 0.6$--$1.0 \times10^{11}\Msun$ plus a comparatively
light DM halo.  \citet{jiao2023} quote
$2.06^{+0.24}_{-0.13}\times10^{11}\Msun$ for their preferred total mass and
a strict upper limit of $5.4\times10^{11}\Msun$.

\begin{figure}[htbp]
\centering
\includegraphics[width=\linewidth,keepaspectratio]{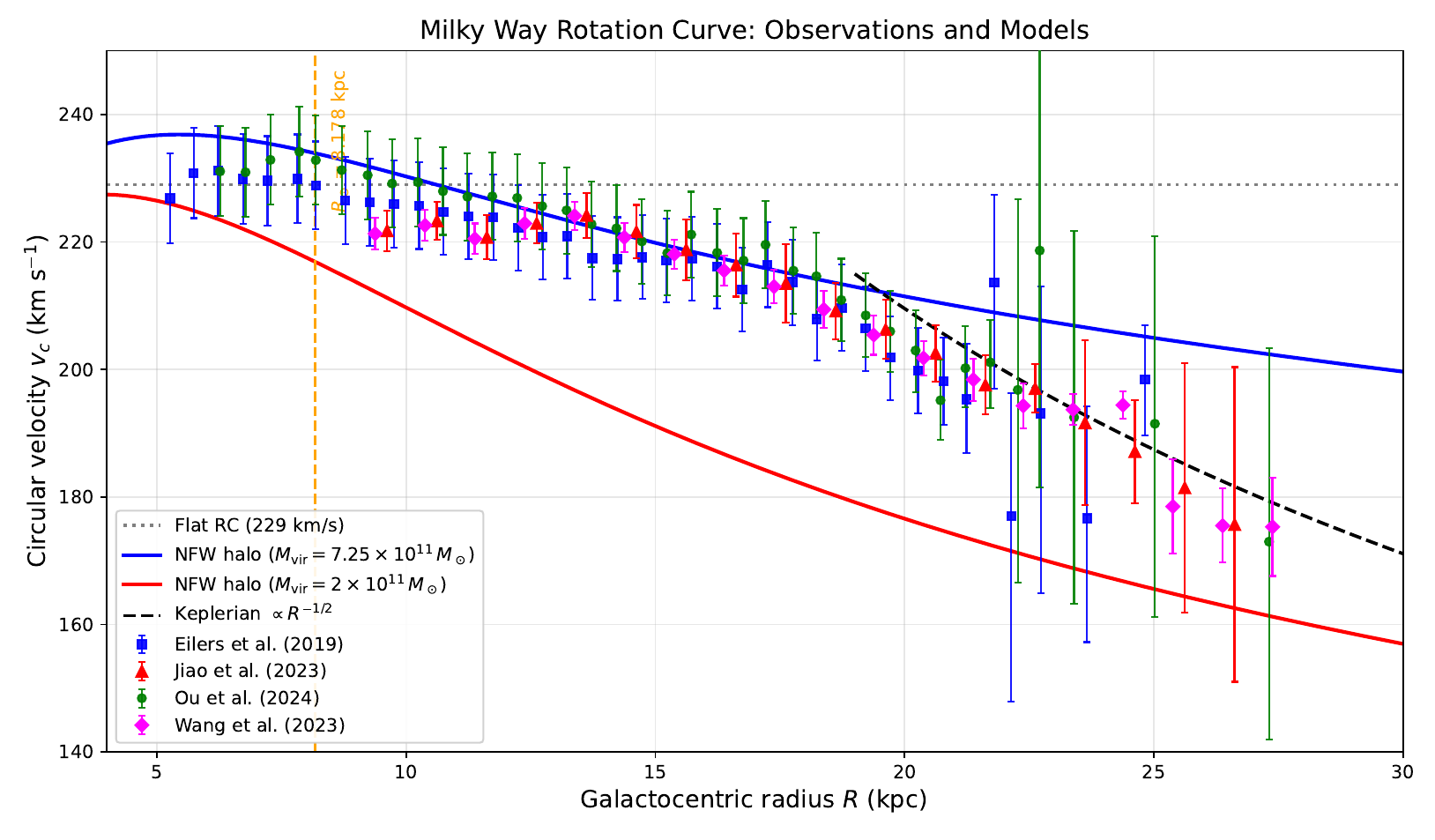}
\caption{MW rotation curve determinations from \textit{Gaia} DR3.
  Data points from \citet{eilers2019} (blue squares), \citet{jiao2023} (red
  triangles), \citet{ou2024} (green circles), and \citet{wang2023} (magenta
  diamonds) are plotted at their published radii and values (E19 Table~1,
  J23 Table~3, O24 Table~1, W23 Table~1).  Error bars: E19 and O24
  bootstrap errors with the authors' systematic budgets added in quadrature
  ($3\%$ for E19; $3\%$ inside $22\kpc$ rising to $15\%$ beyond for O24);
  J23 as published (systematics included); W23 statistical only.  The
  analyses adopt $\Rsun$ between $8.12$ and $8.34\kpc$, so radii are not on
  a common scale at the $\sim0.2\kpc$ level.
  Note that these four determinations use different tracer populations
  (RGB stars in \citeauthor{eilers2019}/\citeauthor{ou2024}, Lucy-inversion
  of the full Gaia DR3 catalogue in \citeauthor{wang2023}, combined reanalysis
  in \citeauthor{jiao2023}) and different treatments of the asymmetric
  drift and distance scale; the points are therefore not independent
  realisations of a single measurement, and the residual systematic
  offsets between them (of order $5$--$10\kms$) should be borne in
  mind when comparing to the model curves.  The outermost
  \citeauthor{eilers2019} bins at $R \gtrsim 18\kpc$ lie below their own
  NFW fit and are consistent, within their larger error bars, with the DR3
  determinations.  Model
  curves show the high-mass NFW profile
  ($M_{\rm vir} = 7.25 \times 10^{11}\Msun$, blue solid), and an
  illustrative light NFW profile ($M_{\rm vir} = 2.06 \times 10^{11}\Msun$,
  red solid).  The latter uses the numerical value of the Jiao+23
  spherical-equivalent total mass only for illustration and is not their
  fitted NFW virial mass.  Also shown are a
  Keplerian decline from
  $R = 19\kpc$ (black dashed), and a flat RC (grey dotted).  The orange
  vertical line marks $\Rsun = 8.178\kpc$.}
\label{fig:rc_data}
\end{figure}

Figure~\ref{fig:rc_data} compares these source measurements with several
illustrative mass-model curves.

\subsection{Ou et al.\ (2024)}
\label{sec:ou2024}

\citet{ou2024} used an axisymmetric Jeans analysis but neglected the
vertical cross-term in equation~\eqref{eq:va2_full}, estimating its omission
as one contribution to their systematic error budget.  Their quoted
systematic uncertainties remain small out to about $20\kpc$ and grow
rapidly beyond $22\kpc$, where the asymmetric-drift contribution alone
reaches the $\sim 15\%$ level \citep{jiao2023}, so the outermost bins
carry substantially larger uncertainties than the inner ones.  Their RC
spans $R = 6.3$--$27.3\kpc$, declining smoothly from
$\vc \approx 231$--$234\kms$ at $6$--$8\kpc$ to $\approx 173\kms$ at
$27.3\kpc$.  Their preferred Einasto fit has
$M_{200}=1.81^{+0.06}_{-0.05}\times10^{11}\Msun$ and
$\rho_{\rm DM,\odot}=0.447\pm0.004\,\mathrm{GeV\,cm^{-3}}$.
The much poorer gNFW fit gives $M_{200}=6.94\pm0.12\times10^{11}\Msun$ and
$\rho_{\rm DM,\odot}=0.405\pm0.004\,\mathrm{GeV\,cm^{-3}}$.

\subsection{Three-dimensional constraints on dark matter geometry}
\label{sec:sylos2023}

Building on \citet{sylos2023}, \citet{syloslabini2024} fit both NFW and a
dark matter disc (DMD) model to the combined \citeauthor{eilers2019} +
\citeauthor{wang2023} RC:
\begin{align*}
  \Mvir^{\mathrm{NFW}} &= (6.5 \pm 0.5) \times 10^{11}\Msun,\quad
    \Rvir = (180 \pm 3)\kpc,\\
  M_{\mathrm{DMD}} &= (1.7 \pm 0.2) \times 10^{11}\Msun.
\end{align*}
The DMD model, in which DM follows an exponential surface-density profile
co-planar with the baryons (cf.\ equation~\ref{eq:vc_expdisk}), fits the
in-plane RC as well as NFW but predicts a very different off-plane velocity
structure---a potentially discriminating test.

That test has now been attempted by
\citet{syloslabini_capuzzo2026}.  Using roughly $1.6\times10^6$ Gaia DR3
stars with full six-dimensional phase-space information, they reconstruct
both a generalized off-plane rotation curve, defined by
$v_c^2(R,z)=R\,\partial\Phi/\partial R$, and the vertical acceleration
$a_z(R,z)$ over $8.5<R/\kpc<14$ and $|z|<2\kpc$.  At $z\ne0$ the former is a
proxy for the radial gravitational field, rather than the speed of a
strictly circular orbit.  The measured velocity components have quoted
errors below about $5\%$, whereas the authors estimate that the reconstructed
vertical acceleration carries model-dependent uncertainties as large as
$\sim20\%$.

The paper compares the same fixed stellar and gaseous components embedded
either in a spherical NFW halo or in a co-rotating exponential dark matter
disc (DMD).  For the fiducial constant tangential-velocity uncertainty
$\Delta v_\phi=3\kms$, the reported reduced $\chi^2$ values for the
off-plane RCs are $1.3$ (NFW) and $0.8$ (DMD).  The corresponding best fits
are
\begin{align*}
  M_{\rm DMD} &= 1.0\times10^{11}\Msun, & R_{\rm DMD}&=7.0\kpc,\\
  M_{\rm vir}^{\rm NFW} &= 8.0\times10^{11}\Msun, & r_s&=18\kpc,
\end{align*}
with total Galactic masses quoted as $1.8\times10^{11}\Msun$ and
$8.8\times10^{11}\Msun$, respectively.  Evaluated at those RC best fits,
the vertical-acceleration comparison also favours the DMD model
($\chi^2=1.8$ rather than $3.3$).  The result is valuable because it adds a
geometric discriminator that is largely absent from midplane RC fits.
However, it does not independently confirm the Keplerian tail: the sample
ends at $14\kpc$, just inside the radius at which the steep outer decline is
usually claimed.  Moreover, the preference for the DMD is conditional on
the effective Jeans closure, tracer scale heights, neglected tilt term, and
restricted pair of mass models discussed in
Section~\ref{sec:offplane_geometry}.

\subsection{A phenomenological MCMC analysis and Gaussian Process reconstruction}
\label{sec:mcmc_gp}

To complement the physically motivated NFW and gNFW fits described in
Section~\ref{sec:mass}, we present an illustrative reanalysis based on a
phenomenological model and, separately, on a Gaussian Process (GP)
reconstruction.  Neither method assumes a parametric DM halo profile, but
both remain conditional on the heterogeneous input compilation, its
covariance treatment, and the adopted regression priors.

\subsubsection{Observational dataset}
\label{sec:dataset}

The analysis is based on the heterogeneous compilation of MW circular-speed
estimates listed in Table~\ref{tab:rc_data}.  It combines disc-tracer RCs
with halo-tracer estimates after rescaling to the common convention
$\Rsun = 8.178\kpc$.
The sample spans $6.3 \le R/\kpc \le 49.0$ and includes 35 measurements
with individual uncertainties $\sigma_{V_c}$ accounting for both statistical
and systematic contributions as reported in the source analyses.  The
points beyond the stellar disc are model-derived circular-speed constraints,
not direct measurements of disc rotation; source covariances are unavailable,
so the 35 entries must not be interpreted as 35 independent observations.

\begin{table}[htbp]
\centering
\caption{Milky Way circular velocity measurements used in the MCMC and
  Gaussian Process analysis. The compilation draws on disc-tracer rotation
  curves from \citet{eilers2019,mroz2019,ablimit2020,zhou2023,wang2023,
  jiao2023,ou2024,syloslabini2024} and halo-tracer determinations from
  \citet{huang2016,ablimit2017,kafle2012,wegg2019}, after correction to a
  common $\Rsun = 8.178\kpc$ scale.  The outer halo-tracer entries are
  model-dependent estimates.  Where multiple authors report values in
  overlapping radial bins, we adopt an inverse-variance weighted mean,
  although unpublished cross-covariances prevent a formally optimal
  combination.}
\label{tab:rc_data}
\begin{tabular*}{\linewidth}{@{\extracolsep{\fill}}ccc|ccc}
\toprule
$R$ [kpc] & $V_c$ [km/s] & $\sigma_{V_c}$ [km/s] &
$R$ [kpc] & $V_c$ [km/s] & $\sigma_{V_c}$ [km/s] \\
\midrule
  6.3 & 217.5 & 22.9 & 21.0 & 206.7 & 17.4 \\
  7.4 & 215.4 & 26.7 & 21.5 & 197.6 &  4.6 \\
  8.9 & 223.1 & 20.8 & 22.4 & 204.7 & 17.7 \\
  9.5 & 221.8 &  3.2 & 22.5 & 197.0 &  3.8 \\
 10.5 & 223.3 &  3.0 & 23.5 & 191.6 & 13.0 \\
 10.9 & 214.1 & 29.5 & 24.5 & 187.1 &  8.1 \\
 11.5 & 220.7 &  3.5 & 25.5 & 181.4 & 19.6 \\
 11.6 & 215.5 & 21.5 & 26.5 & 175.7 & 24.7 \\
 12.5 & 222.9 &  3.2 & 26.9 & 204.1 & 16.7 \\
 13.5 & 224.2 &  3.5 & 28.7 & 203.4 & 14.6 \\
 14.5 & 221.6 &  4.2 & 32.3 & 181.3 & 15.3 \\
 15.5 & 218.8 &  4.8 & 38.2 & 218.9 & 12.8 \\
 16.2 & 205.2 & 19.8 & 39.0 & 196.7 & 14.2 \\
 16.5 & 216.4 &  5.0 & 39.4 & 216.2 & 16.3 \\
 17.5 & 213.5 &  6.1 & 45.5 & 188.6 & 16.3 \\
 18.5 & 209.2 &  4.4 & 49.0 & 172.0 & 12.5 \\
 18.9 & 207.4 & 16.3 &      &       &       \\
 19.5 & 206.2 &  4.6 &      &       &       \\
 20.5 & 202.5 &  4.4 &      &       &       \\
\bottomrule
\end{tabular*}
\end{table}

\subsubsection{Phenomenological fit and MCMC parameter inference}
\label{sec:phenom_fit}

To capture the smooth variation of $\vc(R)$ across both the inner and outer
disc without imposing the specific functional form of an NFW or Einasto
profile, we adopt the monotonic phenomenological model
\begin{equation}
  V(R) = A + \frac{B}{1 + \left(R/C\right)^{n}},
  \label{eq:phenom_vc}
\end{equation}
where $A$, $B$, $C$, and $n$ are free parameters.  The model describes a
smoothed step: an inner plateau at $V(R \to 0) = A + B$, a transition of
characteristic radius $C$ and sharpness $n$, and a monotonic approach to
the outer asymptote $V(R \to \infty) = A$.  The large best-fit value of the
exponent $n$ (see below) produces a sharp, quasi-step transition that
captures the characteristic ``shoulder'' in the data at
$R \approx 18$--$20\kpc$ without forcing a fixed scale radius as in the NFW
parametrisation.  Equation~\eqref{eq:phenom_vc} imposes a monotonic decline
and a constant outer asymptote by construction.  It can describe declines of
different sharpness over the fitted interval, but it is not agnostic about
monotonicity and should not be used to decide whether the true asymptotic
curve is Keplerian.

\begin{figure}[htbp]
\centering
\includegraphics[width=0.5\linewidth,keepaspectratio]{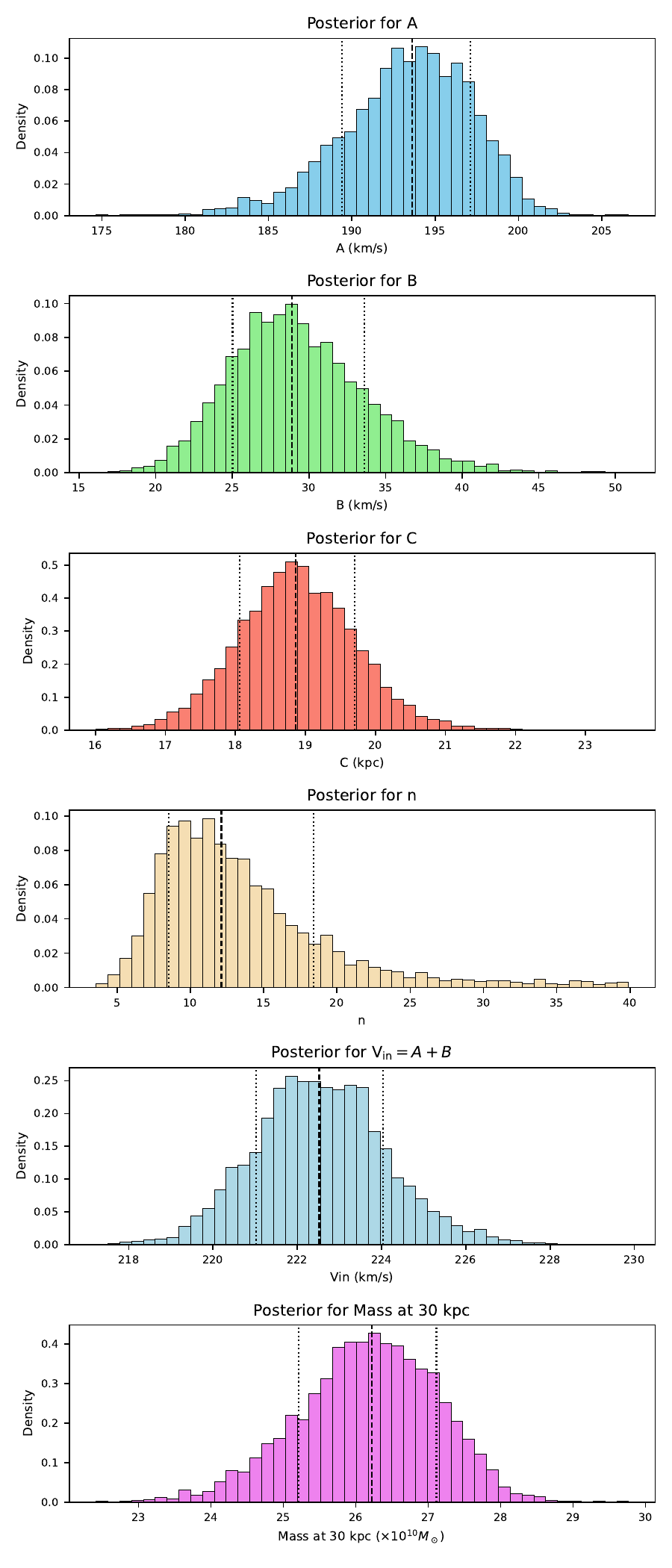}
\caption{Posterior distributions from the MCMC analysis of the phenomenological
  model~\eqref{eq:phenom_vc} applied to the data in Table~\ref{tab:rc_data}.
  The panels show marginalised posteriors for the four model parameters
  ($A$, $B$, $C$, $n$), the inner-plateau velocity $V_{\rm in} = A+B$, and
  the spherical-equivalent mass within $30\kpc$, $M_{\rm sph,30}$.  Vertical
  dashed lines indicate the median and $68\%$ credible intervals.  The
  inferred $M_{\rm sph,30} = 2.62^{+0.09}_{-0.10}\times 10^{11}\Msun$
  is $\sim 25\%$ lower than the mass expected from a flat
  RC at the same radius.}
\label{fig:posterior}
\end{figure}

We first perform a least-squares minimisation against the data in
Table~\ref{tab:rc_data}, finding
$\chi^2 = 14.6$ with $35 - 4 = 31$ degrees of freedom
($\chi^2/\nu \approx 0.47$, p-value $\approx 0.99$).
We then run a full Bayesian inference using
the Metropolis-Hastings MCMC algorithm with $2\times 10^5$ iterations
(first $10\%$ discarded as burn-in, thinning interval of 50).
The intervals below are conditional on the adopted likelihood, parameter
priors, and diagonal covariance model.  The numerical implementation,
including the priors, proposal settings, chains, and convergence outputs, is
available from the authors as described in the Data and Code Availability
statement.
The posterior distributions for all parameters and the
derived quantities are shown in Figure~\ref{fig:posterior}.

The 68\% credible intervals for the best-fit parameters are:
$A = 193.6^{+3.5}_{-4.2}\kms$,
$B = 28.9^{+4.8}_{-3.9}\kms$,
$C = 18.9^{+0.8}_{-0.8}\kpc$, and
$n = 12^{+6}_{-4}$.
The derived inner-plateau velocity, defined as the inner asymptotic value
$V_{\rm in} = V(R\to 0) = A + B$, is inferred as
$V_{\rm in} = 222.5^{+1.5}_{-1.5}\kms$, somewhat below the canonical
$\vc(R_0) \approx 229\kms$ of \citet{eilers2019}; this reflects the weight
of the \citet{jiao2023} points in the compilation, which lie
$\sim5\kms$ below the RGB-based curves at $R<13\kpc$ and are referred to
$\Rsun=8.34\kpc$.  The outer asymptote is
$V_{\rm out} = A = 193.6^{+3.5}_{-4.2}\kms$.  We stress that these are
properties of the fitting function within and slightly beyond the radial
range of the data; the phenomenological model is not intended to be
extrapolated to arbitrarily large radii.

\begin{figure}[htbp]
\centering
\includegraphics[width=\linewidth,keepaspectratio]{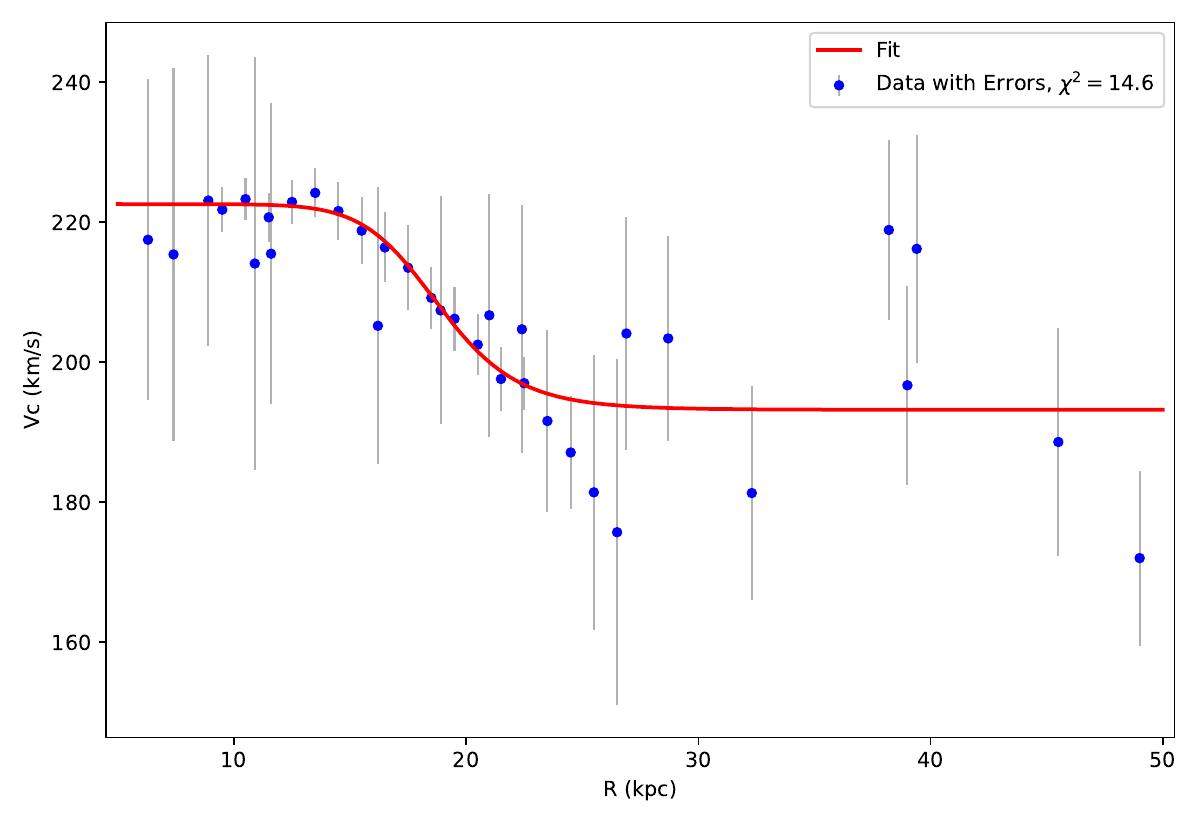}
\caption{Best-fit phenomenological model~\eqref{eq:phenom_vc} (solid curve)
  to the MW rotation curve data of Table~\ref{tab:rc_data} (points with error
  bars).  The reduced chi-square of the fit is
  $\chi^2/\nu = 14.6/31 \approx 0.47$; see the text for a discussion of
  the statistical interpretation of this low value.}
\label{fig:rc_fit_phenom}
\end{figure}

Figure~\ref{fig:rc_fit_phenom} shows the best-fit phenomenological curve
against the data.  The fit qualitatively reproduces all the features visible
in the data: the nearly flat plateau at $\sim220$--$225\kms$ for
$R < 15\kpc$, the
transition region at $R \approx 15$--$20\kpc$, and the declining tail at
$R > 20\kpc$.

The low goodness-of-fit statistic requires care.  A reduced chi-square
substantially below unity ($\chi^2/\nu \approx 0.47$ here)
should \textit{not} be over-interpreted as evidence of an exceptionally
good model.  Values of $\chi^2/\nu \ll 1$ typically signal that the quoted
uncertainties are conservative (systematic error budgets added in
quadrature to statistical ones), that the measurements are not mutually
independent, or that positive correlations between data points have been
neglected.  All three effects are plausibly at work in
Table~\ref{tab:rc_data}: the compilation combines determinations obtained
with different tracers but partially overlapping stellar samples and
common calibrations (distance scale, solar motion, asymmetric drift
treatment), and adjacent radial bins within a single analysis are known to
be correlated (Section~\ref{sec:correlations}; \citealt{oman2024}).  The
diagonal-covariance likelihood adopted here therefore very likely
overestimates the effective number of independent constraints, and the
credible intervals quoted in this section should be regarded as approximate.  A full
treatment would require the (unpublished) cross-covariances between the
input analyses.

\subsubsection{Enclosed mass and the mass deficit relative to a flat RC}
\label{sec:mass_deficit}

\begin{figure}[htbp]
\centering
\includegraphics[width=0.85\linewidth,keepaspectratio]{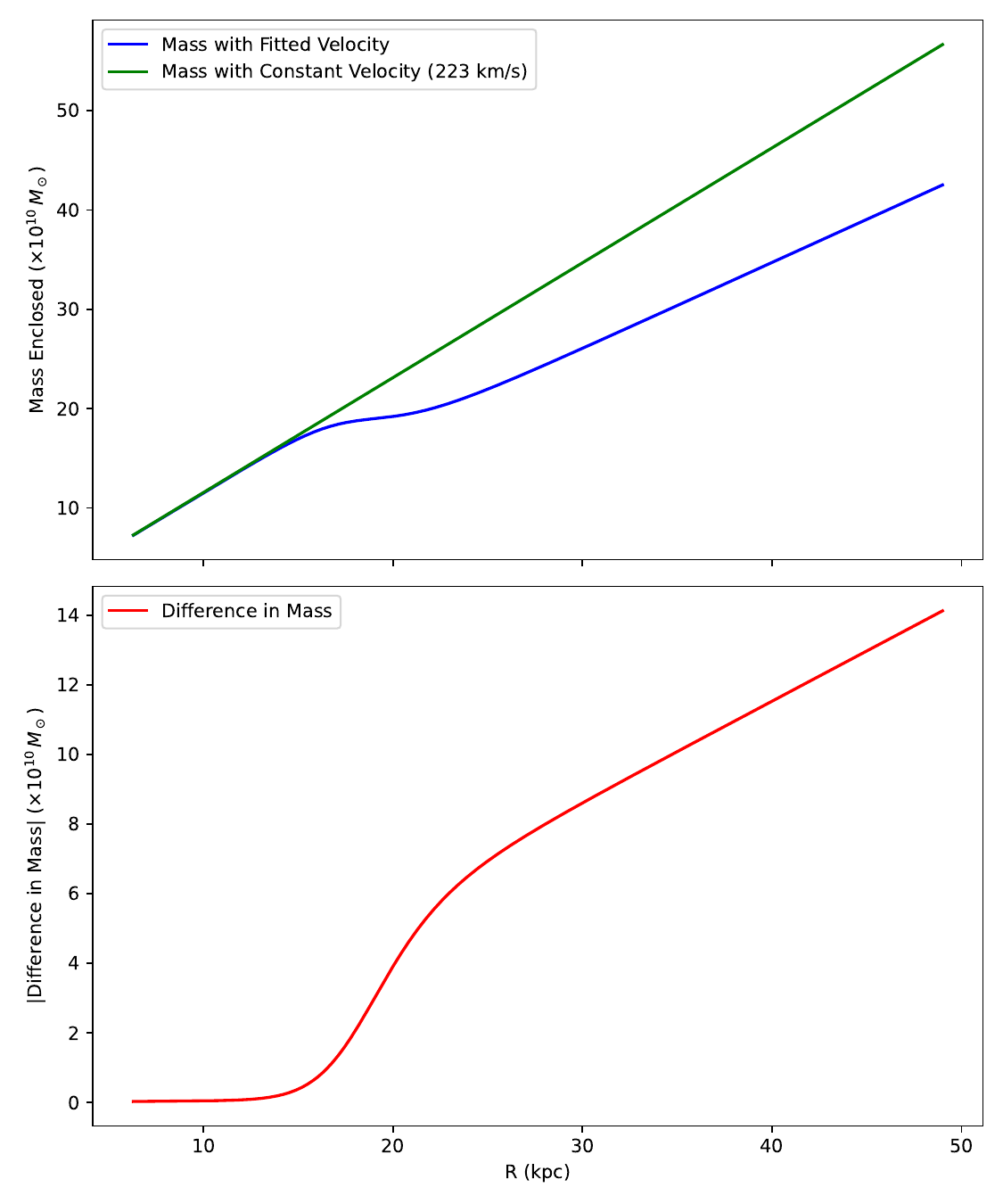}
\caption{Top panel: spherical-equivalent mass $M_{\rm sph}(<R)$ computed from the phenomenological
  best-fit model~\eqref{eq:phenom_vc} (blue curve) and
  from a constant circular velocity $\vc = 223\kms$ (green solid line).
  Both curves agree well at $R \lesssim 15\kpc$ but diverge significantly
  beyond this radius.  Bottom panel: absolute mass difference $\Delta M(R)$
  between the two spherical-equivalent curves, which grows monotonically beyond the transition
  radius and reaches $\Delta M \approx 8.5 \times 10^{10}\Msun$ at
  $R = 30\kpc$; it quantifies the difference relative to a canonical
  flat-RC extrapolation, not mass physically removed from the Galaxy.}
\label{fig:mass_plots}
\end{figure}

The spherical-equivalent dynamical mass is computed from
equation~\eqref{eq:vc_spherical_mass} using the fitted velocity curve:
$M_{\rm sph}(<R) = V(R)^2 R / G$.  It equals the true enclosed mass only
under spherical symmetry.  Figure~\ref{fig:mass_plots} compares
$M_{\rm sph}(<R)$ from the declining fit with the corresponding value
inferred from a \textit{flat} RC at $\vc = 223\kms$.

Within the spherical conversion, the mass within $30\kpc$ from the MCMC
analysis is $M_{\rm sph,30} = 2.62^{+0.09}_{-0.10}\times 10^{11}\Msun$,
while the flat-RC estimate gives
$M_{\rm sph,30}^{\rm flat} \approx 3.5\times 10^{11}\Msun$.  The
discrepancy at $30\kpc$ is $\Delta M \approx 8.5^{+1.0}_{-0.9}\times
10^{10}\Msun$ (68\% credible interval).  Because the fitted velocity
approaches a constant outer asymptote below the flat-RC value, $\Delta
M(R)$ continues to grow with radius, reaching $\approx 1.4 \times
10^{11}\Msun$ at the outermost data point ($R = 49\kpc$); the value at
$30\kpc$ is quoted as a representative figure within the radial range
where the data are densest.
This difference cannot be compared directly with the present bound mass of
the Sagittarius dwarf: a perturbation can bias an equilibrium kinematic
inference without supplying or removing an equal amount of gravitating mass.

\subsubsection{Gaussian Process reconstruction}
\label{sec:gp}

To test sensitivity to the specific functional form
adopted in equation~\eqref{eq:phenom_vc}, we also reconstructed $\vc(R)$
using \textit{Gaussian Processes} (GPs), a fully non-parametric Bayesian
regression framework \citep{rasmussen2006}.  GPs assign a prior
probability distribution over functions; the posterior distribution,
conditioned on the observational data, represents the family of smooth
curves consistent with the measurements.

A Gaussian process is characterised by its mean function $\mu(x)$ and
covariance kernel $k(x, x')$ (not to be confused with the velocity-ellipsoid
ratio $\kappa^2$ of Section~\ref{sec:asym_drift}):
\begin{equation}
  f(x) \sim \mathcal{GP}\!\bigl(\mu(x),\,k(x, x')\bigr).
  \label{eq:gp_def}
\end{equation}
We adopt a zero prior mean and the squared-exponential (SE) kernel
\begin{equation}
  k_{\rm SE}(r) = \sigma_f^2\exp\!\left(-\frac{r^2}{2l_f^2}\right),
  \quad r = x - x',
  \label{eq:se_kernel}
\end{equation}
where $\sigma_f$ is the signal amplitude (standard deviation) and $l_f$ is
the correlation length.
The hyperparameters $(\sigma_f, l_f)$ are optimised by maximising the log
marginal likelihood:
\begin{equation}
  \ln \mathcal{L}(\Theta)
  = -\tfrac{1}{2}\bm{y}^T K^{-1}\bm{y}
    - \tfrac{1}{2}\ln|K|
    - \tfrac{n}{2}\ln(2\pi),
  \label{eq:gp_loglike}
\end{equation}
where $K = k(x_i, x_j) + \Sigma_{n\times n}$ is the sum of the kernel
matrix and the observational noise covariance (diagonal, with entries
$\sigma_{V_c}^2$).  Marginalising over the function values at unobserved
points gives the predictive Gaussian distribution, from which we extract
the mean and $1\sigma$ ($2\sigma$, $3\sigma$) posterior credible bands.  The
analysis was performed with the GaPP code \citep{seikel2012}.

\begin{figure}[htbp]
\centering
\includegraphics[width=0.85\linewidth,keepaspectratio]{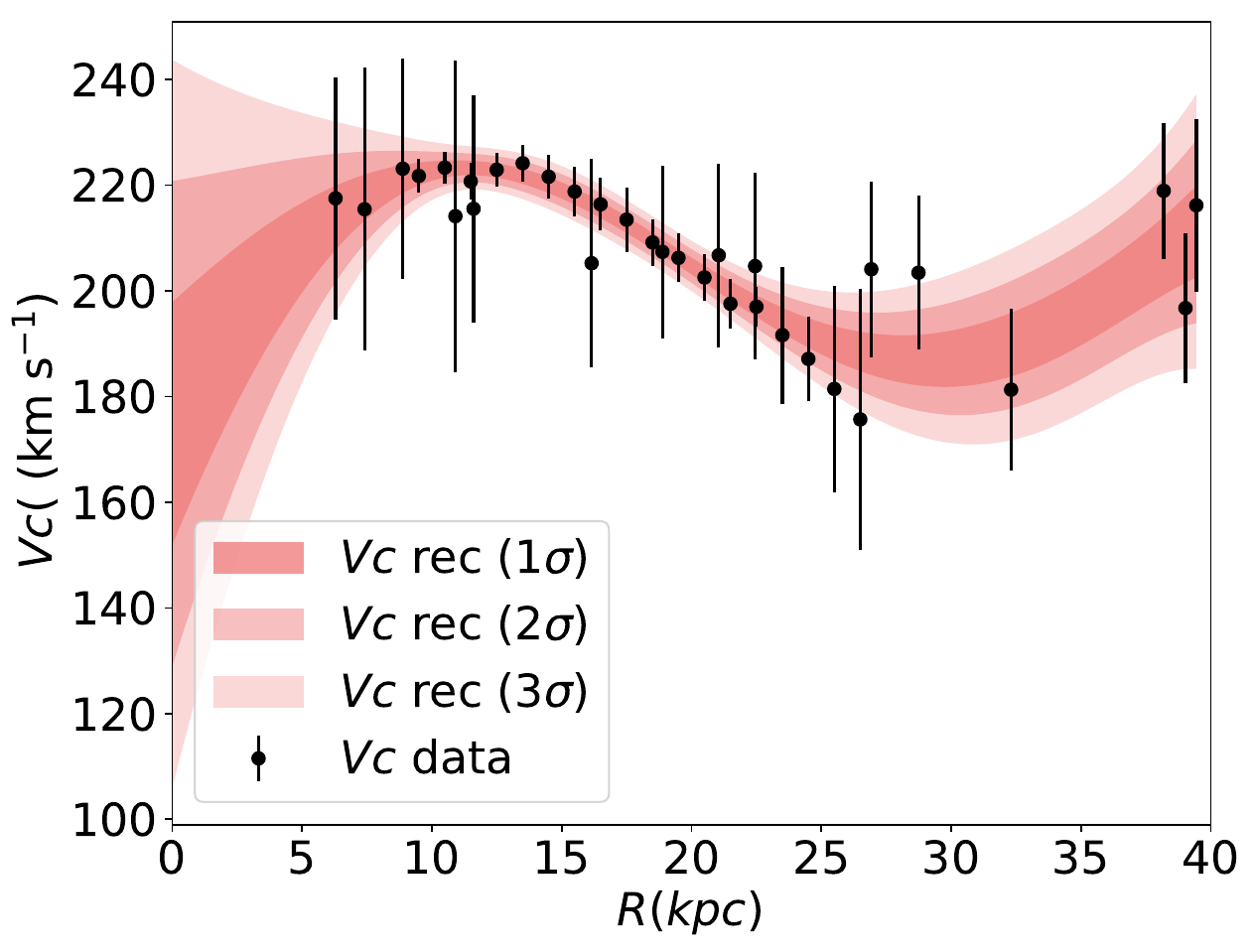}
\caption{Gaussian Process reconstruction of the MW circular velocity.  The
  posterior mean (thick curve) with $1\sigma$, $2\sigma$, and $3\sigma$
  posterior credible bands (shaded salmon regions) is shown together with the
  observational data (black points with error bars) from
  Table~\ref{tab:rc_data}.  The SE kernel hyperparameters are optimised
  by maximising the log marginal likelihood~\eqref{eq:gp_loglike}.  The
  decline is conditional on the input compilation, kernel, and diagonal
  observational covariance.}
\label{fig:gp_vc}
\end{figure}

Figure~\ref{fig:gp_vc} shows the GP reconstruction of $\vc(R)$.  The
posterior mean is consistent with the phenomenological fit of
equation~\eqref{eq:phenom_vc}, showing a flat inner plateau and a lower mean
velocity beyond $20\kpc$ under the adopted diagonal noise model.  A unique
statistical significance for the decline is not quoted because it depends on
the radii being contrasted and on the unmodelled source covariance.  The GP does not impose
a parametric halo profile, but its result depends on the zero mean,
squared-exponential kernel, optimised hyperparameters, and the neglected
source covariances.

\begin{figure}[htbp]
\centering
\includegraphics[width=0.8\linewidth,keepaspectratio]{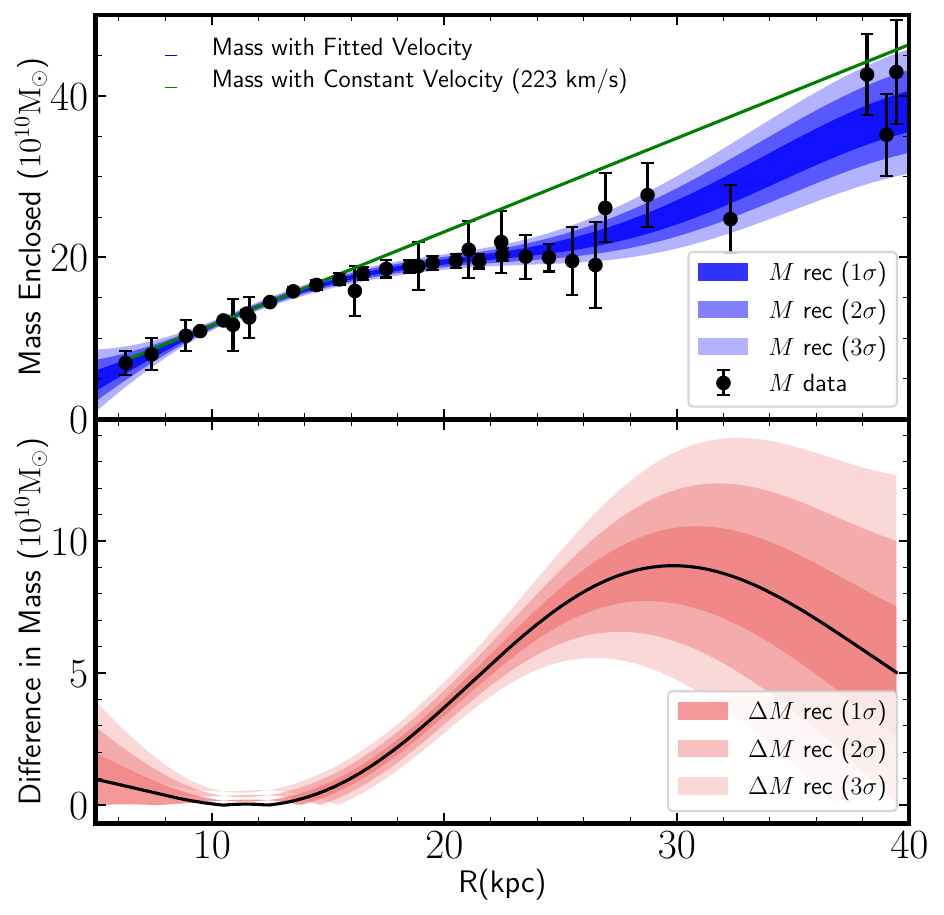}
\caption{GP-based spherical-equivalent mass reconstruction.
  \textbf{Top:} Posterior mean and $3\sigma$ bands (salmon) for
  $M_{\rm sph}(<R)$ from the GP reconstruction
  of $\vc(R)$, compared with the flat-RC mass (green dashed, assuming
  $\vc = 223\kms$).  The two profiles agree for $R \lesssim 15\kpc$ and
  diverge thereafter.  \textbf{Bottom:} Absolute mass difference $\Delta M(R)$
  between the GP reconstruction and the flat-RC baseline, with the posterior
  mean (black line) and $3\sigma$ band (salmon).  The mass deficit peaks at
  $\Delta M_{\rm sph} \approx 8$--$9 \times 10^{10}\Msun$ near
  $R \approx 30\kpc$.}
\label{fig:gp_mass}
\end{figure}

Figure~\ref{fig:gp_mass} shows the GP-inferred spherical-equivalent mass profile and
its difference from the flat-RC baseline, confirming the result from the
phenomenological fit: a mass deficit of $\Delta M \sim 8$--$9 \times
10^{10}\Msun$ builds up for $R \gtrsim 20\kpc$ and peaks at $R \approx
30\kpc$.  The agreement between the MCMC fit and the GP reconstruction shows
that the feature is not unique to equation~\eqref{eq:phenom_vc}; because the
methods use the same compilation and diagonal noise model, it is not an
independent confirmation of the feature's physical origin.

The phenomenological MCMC and GP analyses thus provide a
non-halo-parametric consistency check on the RC decline first reported by
\citet{eilers2019} and subsequently strengthened by \citet{jiao2023} and
\citet{ou2024}.  The inferred spherical-equivalent mass within $30\kpc$,
$M_{\rm sph,30} \approx 2.6\times 10^{11}\Msun$, exceeds the strict
Keplerian spherical-equivalent total of $2.06\times 10^{11}\Msun$
\citep{jiao2023}.  It cannot be ranked directly against NFW virial-mass
estimates of $(5$--$8)\times 10^{11}\Msun$ \citep{eilers2019,ou2024}, because
those refer to a different radius and depend on halo extrapolation.  Our
phenomenological model evaluates only the spherical-equivalent relation over
the data range and does not extrapolate to a virial radius.  Neither analysis
assumes a DM halo profile, concentration, or radial extent, but neither is
statistically independent of the source data or free of regression and
covariance assumptions.

\subsection{Summary of RC measurements}

Table~\ref{tab:rc_summary} summarises the principal recent determinations and
keeps enclosed-mass estimates distinct from virial masses.

\begin{table}[htbp]
\centering
\caption{Recent MW rotation curve determinations.  The outer-disc slope
  $\dif\vc/\dif R$ is measured over $R \approx 15$--$25\kpc$.  The
  Cepheid-based analysis of \citet{feng2026}, which finds only a mild
  decline over $6$--$18\kpc$, is included for comparison.  LIM: Lucy
  inversion method.  The slopes for Wang+23 and Ou+24 are the linear fits
  quoted by \citet{wang2023} and \citet{jiao2023}, respectively.  For
  \citet{jiao2023} we quote the linear slope; their separate power-law
  fit beyond $19\kpc$ gives $\gamma = -0.47 \pm 0.15$, consistent with
  the Keplerian value $-0.5$ (Section~\ref{sec:jiao2023}).  The
  Sylos Labini--Capuzzo-Dolcetta 2026 entry is instead a three-dimensional
  force-field analysis; its two masses refer to the fitted DMD mass and NFW
  virial mass and are not like-for-like estimates.}
\label{tab:rc_summary}
\scriptsize
\resizebox{\linewidth}{!}{%
\begin{tabular}{lcccc}
\toprule
Study & Tracer & $R$ range & Slope & Mass estimate $(10^{11}\Msun)$\\
\midrule
Eilers+19       & RGB (APOGEE)   & 5--25 kpc   & $-1.7\pm0.1$   & $7.25\pm0.26$ \\
Mróz+19         & Cepheids       & 4--20 kpc   & $-1.34\pm0.21$ & ---           \\
Wang+23         & All Gaia DR3 (LIM) & 9.5--27.5 kpc & $-2.3\pm0.2$ & ---           \\
Sylos L.+23     & E19+W23        & 5--28 kpc   & declining      & $6.5\pm0.5$   \\
Jiao+23         & W23 (re-derived)+O24 & 9.5--26.5 kpc & $-2.18\pm0.23$ & $M_{\rm tot}=2.06^{+0.24}_{-0.13}$\\
Ou+24           & RGB (APOGEE)   & 6.3--27.3 kpc & $-2.22\pm0.20$ & $M_{200}=1.81$ (Einasto)  \\
Sylos L.\&C.-D.+26 & Gaia DR3 6D & 8.5--14 kpc & 3D force field & DMD: $1.0$; NFW: $8.0$ \\
Feng+26         & Cepheids (Gaia DR3) & 6--18 kpc & mild         & ---    \\
\bottomrule
\end{tabular}%
}
\end{table}

\section{Mass Models and the Consequences of a Keplerian Decline}
\label{sec:mass}

\subsection{NFW mass inference procedure}

Given a set of measured RC values $\{\vc^{\rm obs}(R_i), \sigma_i\}$, we fit
the model $\vc^2 = v_{c,\rm bary}^2 + v_{c,\rm NFW}^2$ via:
\begin{equation}
  \chi^2 = \sum_{i=1}^N
    \frac{\bigl[\vc^{\rm obs}(R_i) - \vc^{\rm model}(R_i; r_s, \rho_s)\bigr]^2}
         {\sigma_i^2}.
  \label{eq:chi2}
\end{equation}
The baryonic contribution uses equations~\eqref{eq:vc_bulge},
\eqref{eq:vc_mn}, and \eqref{eq:vc_expdisk}; the DM part uses
equation~\eqref{eq:nfw_vc}.  A Bayesian approach adopts a concentration-mass
prior from $\Lambda$CDM simulations \citep{dutton2014,bullock2001} and
marginalises over baryonic parameters.

The critical limitation is the extrapolation required to obtain the virial
mass.  The RC is measured only to $R_{\rm max} \approx 26$--$28\kpc$, which
is approximately $15\%$ of $R_{200} \approx 180$--$200\kpc$.  Converting the
observed RC over this limited range to a total virial mass requires assuming
a specific DM density profile over the remaining $85\%$ of the radial extent.  The disc RC probes only the inner halo, and the leverage of a
declining RC measured over $\sim 15$--$28\kpc$ on the \textit{total}
Galactic mass is correspondingly weak: a low total mass inferred from it
is an extrapolation, not a measurement.  Conversely, the main limitation of
the outer-halo estimates is the scarcity of numerous and accurate tracers
beyond $\sim 80\kpc$.  Distant halo stars, globular clusters, satellites and
stellar streams therefore remain indispensable for constraining the total
halo mass, notwithstanding the equilibrium caveats discussed in
Section~\ref{sec:systematics}.

\subsection{Consequences of a genuinely Keplerian decline}

If the decline is genuinely Keplerian ($\vc \propto R^{-1/2}$ for
$R > R_{\rm kep}$), the spherical-equivalent mass
$M_{\rm sph}(<R) = \vc^2 R/G$
is constant for $R > R_{\rm kep} \approx 19\kpc$, implying that the
contribution of mass at $R > 19\kpc$ to the gravitational budget is
negligible.  The total mass then gives
$M_{\rm tot} \approx 2.04 \times 10^{11}\Msun$ from
equation~\eqref{eq:jiao_mass}.  The conclusion is exact only for a spherical
potential and is in strong tension with many outer-halo constraints
(Section~\ref{sec:indep}).

\begin{figure}[htbp]
\centering
\includegraphics[width=\linewidth,keepaspectratio]{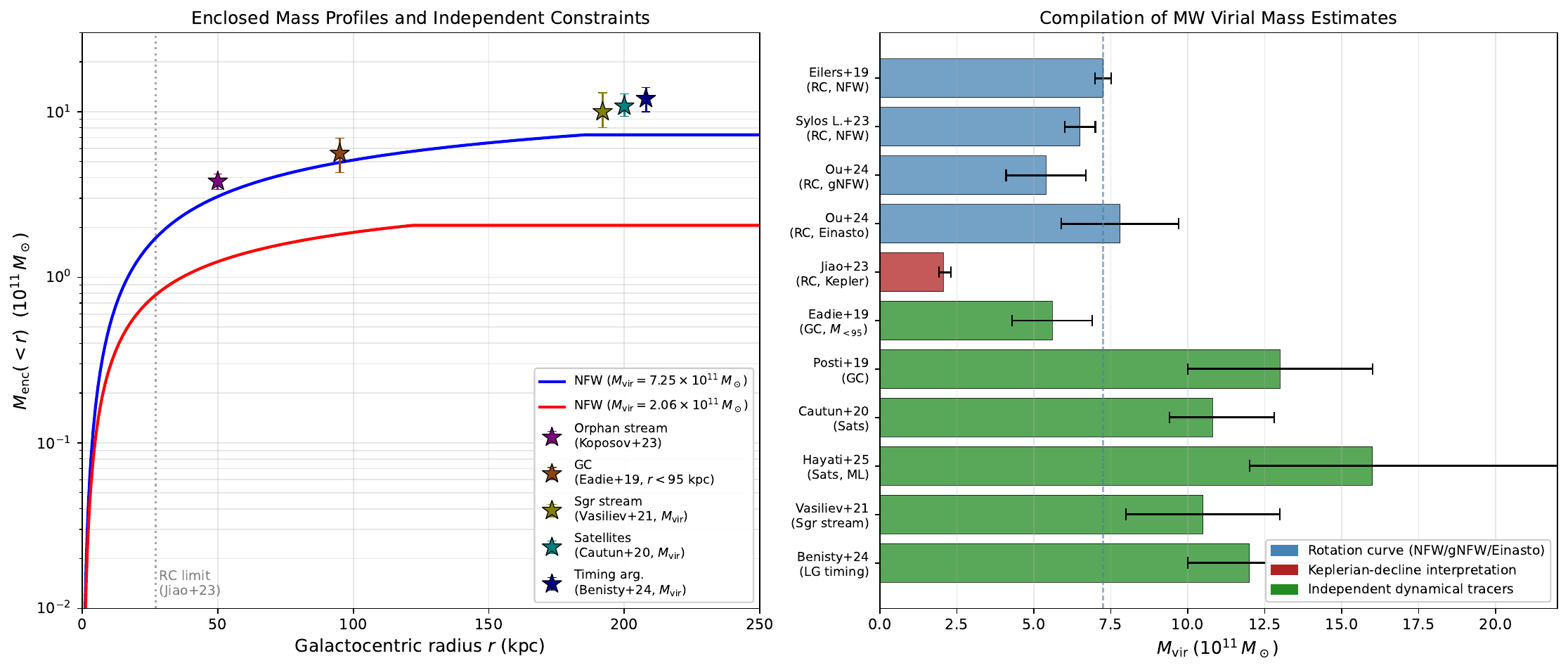}
\caption{\textbf{Left:} Enclosed mass profiles for illustrative NFW models
  with $M_{\rm vir} = 7.25 \times 10^{11}\Msun$ and
  $M_{\rm vir} = 2.06 \times 10^{11}\Msun$ compared with constraints
  from stellar streams (purple, \citealt{koposov2023}), globular clusters
  (brown, \citealt{eadie2019}), satellite kinematics (teal, \citealt{cautun2020}),
  and the timing argument (navy, \citealt{benisty2024}).  The vertical dashed
  line marks the outer limit of the Jiao+23 RC data.  The second NFW curve
  uses the numerical value of the Jiao+23 spherical-equivalent total mass
  only as an illustration; it is not their fitted NFW virial mass.
  \textbf{Right:} Compilation of virial mass estimates: blue/red bars are
  RC-based, green bars are from other dynamical tracers.  The comparison is
  illustrative because the entries constrain different mass definitions and
  adopt different halo families.}
\label{fig:mass_comp}
\end{figure}

Figure~\ref{fig:mass_comp} visualises the radial leverage and the
model-dependent extrapolation behind this tension.

\section{Systematic Uncertainties}
\label{sec:systematics}

\subsection{Overview}

The inference of $\vc(R)$ from the Jeans equation~\eqref{eq:jeans_R} requires
an approximately steady and axisymmetric tracer population with negligible
mean streaming, reliable distances and covariances, and correctly modelled
tracer density and velocity moments (including the radial--vertical tilt
term).  The asymmetric-drift correction is a consequence of those measured
moments rather than a separate dynamical assumption.

\begin{figure}[htbp]
\centering
\includegraphics[width=\linewidth,keepaspectratio]{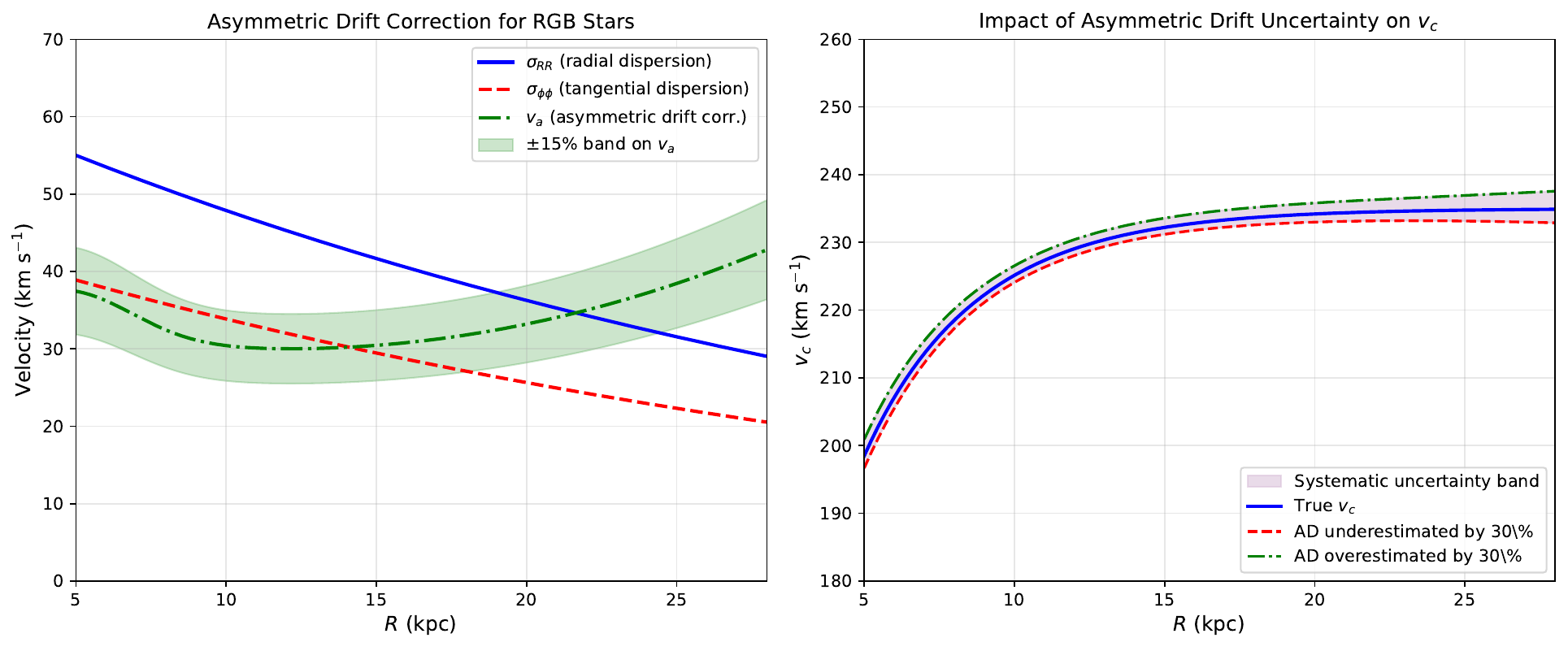}
\caption{\textbf{Left:} Modelled velocity dispersions and resulting asymmetric
  drift correction $v_a$ (equation~\ref{eq:va2_simple}) for an RGB population.
  We adopt $\sigma_{RR}$ declining from $\sim 55\kms$ at $R = 5\kpc$ to
  $\sim 32\kms$ at $R = 25\kpc$ and $\kappa^2 = 0.5$; $v_a$ is at the
  $\sim 30$--$40\kms$ level across the disc for these illustrative tracer
  parameters.
  \textbf{Right:} Impact of a $30\%$ error in the AD correction on the
  inferred $\vc$.  Because $\delta\vc \approx (v_a/\vc)\,\delta v_a$ with
  $v_a/\vc \sim 0.13$, the resulting bias is at the $\sim 1$--$3\kms$
  level at $R > 20\kpc$---small compared to the $\sim 30\kms$ outer
  decline. The asymmetric drift is therefore not, on its own, a dominant
  systematic in the RC inference; the larger budget comes from correlated
  mis-estimation of the tracer profile (which biases both $v_a$ and the
  pressure-support term) together with distance-scale and non-equilibrium
  effects discussed in Section~\ref{sec:systematics}.}
\label{fig:asym_drift}
\end{figure}

Figure~\ref{fig:asym_drift} illustrates the scale of the quadratic
asymmetric-drift term and its local propagation into $\vc$.

\subsection{Asymmetric drift uncertainty}

From equation~\eqref{eq:va2_full}, a conservative first-order absolute-error
bound on $v_a^2$ is
\begin{equation}
  \delta(v_a^2)
  \approx 2\sigma_{RR}\,|\mathcal{A}|\,\delta\sigma_{RR}
          + \sigma_{RR}^2\,\delta|\mathcal{A}|,
  \label{eq:delta_va2}
\end{equation}
where $\mathcal{A} = \pp\ln\nu/\pp\ln R + \pp\ln\sigma_{RR}^2/\pp\ln R
+ 1 - \kappa^2$.  At $R = 20\kpc$ with $\sigma_{RR} \approx 35\kms$,
$|\mathcal{A}| \approx 2.4$ (Section~\ref{sec:asym_drift}),
$\delta\sigma_{RR} = 0.2\,\sigma_{RR}$, and
$\delta|\mathcal{A}| = 0.3\,|\mathcal{A}|$:
\begin{equation}
  \delta(v_a^2) \approx 2\,(35)\,(2.4)\,(7) + (35)^2\,(0.72)
                \approx 2060\,(\kms)^2,\quad
  \delta v_a \approx 19\kms.
\end{equation}
The corresponding shift in the inferred $\vc$ is
$\delta \vc \approx \delta(v_a^2)/(2\vc) \approx 4\kms$, i.e.\ at the
few-per-cent level. While the AD term itself is therefore not the
dominant systematic in the outer-disc RC, the ingredients that enter
$v_a$ ($\sigma_{RR}$, the tracer density profile and its gradient, the
anisotropy $\kappa^2$) also enter the radial pressure-support term in
the Jeans equation and can produce \textit{correlated} biases that
exceed the naive quadrature sum. Simulations from the FIRE-2 suite
(Feedback In Realistic Environments)
\citep{ou2024_fire} find combined biases of $\sim 10$--$40\kms$ at
$R \sim 20$--$25\kpc$ when the full simulation kinematics are passed
through a Jeans-equation pipeline mimicking observational analyses.
A complementary Gaia DR3 Jeans-equation reanalysis by \citet{koop2024},
using both observational data and $N$-body simulations including a
Sagittarius-like perturber, finds that the difference between the true
circular-speed curve and the Jeans-inferred one can reach $\sim 10$--$15\%$
in MW-like systems, and cautions specifically that a radially truncated
tracer density can produce a spuriously steep apparent decline at large
$R$ even when the underlying RC is essentially flat.

\subsection{The choice of Jeans formalism at large radii}
\label{sec:jeans_inconsistency}

\citet{klacka2025} showed that in the best-fit mass models of
\citet{jiao2023} and \citet{ou2024}, the spherically distributed DM halo
accounts for $\gtrsim 88\%$ of the total radial gravitational acceleration at
$R = 27\kpc$, and argued on this basis that the cylindrical Jeans analysis
becomes inappropriate there.  Some care is required in stating this
criticism.  A spherical potential is a special case of an axisymmetric one,
so the axisymmetric Jeans equations in cylindrical coordinates,
equation~\eqref{eq:jeans_R}, remain mathematically valid regardless of
whether the acceleration is dominated by the disc or by a spherical halo:
the appropriate choice of coordinates and Jeans equations is dictated by
the symmetry and phase-space structure of the \textit{tracer population}
and by the closure assumptions, not by the shape of the dominant mass
component.  The legitimate concern raised by this line of argument is
instead the following: the disc-star tracers used at $R \gtrsim 20\kpc$
occupy a flared, warped, and radially truncated distribution whose velocity
ellipsoid orientation (and hence the cross-term
$\sigma_{Rz}^2$) is poorly constrained, and the closure assumptions
(alignment of the ellipsoid with cylindrical or spherical coordinates,
steady state, vanishing mean radial motion) that are benign where the disc
dominates are much less well tested in the halo-dominated regime.  In this
sense the halo dominance at large $R$ does not invalidate
equation~\eqref{eq:jeans_R}, but it does amplify the sensitivity of the
inferred $\vc$ to the adopted tracer geometry and closure assumptions, and
motivates cross-checks with the spherical Jeans
equation~\eqref{eq:jeans_sph} applied to genuinely spheroidal tracers.

An empirical check on the tracer population is, however, already
available.  \citet{ou2024} examine the orbits of their outer-disc sample
individually (their Appendix~A and Figures~A2--A3): of the 244 stars at
$R > 20\kpc$, 241 are found to move on nearly circular, disc-confined
orbits, with only three showing halo-like kinematics.  The outer-disc
tracers therefore appear to be a genuine cold disc population rather than
a mixture contaminated by halo stars, so that the measured RC at these
radii is not obviously altered by halo-related motions.  This weakens the
strong form of the \citeauthor{klacka2025} objection: the issue is not
that the tracers themselves belong to the halo, but the more subtle one,
stated above, of how well the velocity ellipsoid and closure assumptions
of a flared and truncated disc population are constrained where the halo
dominates the acceleration.

\subsection{Off-plane Jeans constraints on dark matter geometry}
\label{sec:offplane_geometry}

The off-plane analysis of \citet{syloslabini_capuzzo2026} illustrates both
the promise and the present limitations of using Gaia to determine the
three-dimensional geometry of the Galactic potential.  For a selected
stellar tracer population, the exact steady, axisymmetric vertical Jeans
equation can be written as
\begin{equation}
  a_z = \frac{1}{\nu}\frac{\partial(\nu\sigma_{zz}^2)}{\partial z}
       +\frac{1}{R\nu}\frac{\partial(R\nu\sigma_{Rz}^2)}{\partial R},
  \qquad a_z\equiv-\frac{\partial\Phi}{\partial z}.
  \label{eq:vertical_force_tracer}
\end{equation}
Here $\nu$ is the number density of the \emph{observed tracer}, whereas the
total mass density enters only by sourcing $\Phi$ through the Poisson
equation.  For an exponential tracer density in $|z|$, the first term gives
\begin{equation}
  a_z\supset
  -\operatorname{sgn}(z)\frac{\sigma_{zz}^2}{h_{\nu,z}},
  \label{eq:vertical_scale_sensitivity}
\end{equation}
so the inferred force is particularly sensitive to the
selection-corrected tracer scale height.

\citet{syloslabini_capuzzo2026} instead adopt an explicitly phenomenological
mean-field closure: observed stellar velocity moments are treated as
effective estimates of the moments of the total gravitating system.  This is
not an exact multicomponent dynamical decomposition, as the authors
themselves emphasize.  Their simplified equations also neglect the tilt
term $\sigma_{Rz}^2$.  In addition, the effective vertical scale entering
the reconstruction changes from $\bar z_d=0.20\pm0.02\kpc$ for the
disc--NFW model to $0.22\pm0.04\kpc$ for the DMD model.  Consequently the
reconstructed $a_z$ data vector is partly conditional on the mass hypothesis
being tested; the quoted $\chi^2$ contrast is therefore not equivalent to a
standard comparison in which the same model-independent measurements and
covariance matrix are held fixed.

The model comparison is also deliberately narrow: a spherical NFW halo is
contrasted with a massive exponential, co-rotating dark disc.  Intermediate
possibilities---an oblate halo, a thick dark disc plus halo, radially varying
tracer scale heights, or a non-equilibrium distribution-function model---are
not fitted.  The assumed constant velocity error controls the absolute fit
quality (the reported reduced $\chi^2$ changes from $26$ to $0.5$ for NFW as
$\Delta v_\phi$ is varied from $1$ to $5\kms$), and correlations between
radial and vertical bins are not included in a full model-selection
calculation.  Finally, evidence for CO-dark molecular gas demonstrates that
some baryonic material is missed by standard tracers, but does not by itself
establish a hidden disc with the required mass
$M_{\rm DMD}\simeq10^{11}\Msun$.

The appropriate interpretation is therefore that the paper identifies a
genuinely useful and testable three-dimensional signature, and reports an
interesting preference within its adopted closure and model family, rather
than a model-independent detection of disc-like dark matter.  A decisive
test should infer the selection-corrected density and velocity ellipsoid of
several stellar populations jointly with the potential, retain the tilt
term, use the full covariance, and compare spherical, oblate, and composite
disc--halo models on the same data vector.

\subsection{Distance calibration systematics}
\label{sec:dist_syst}

Tangential velocities scale linearly with the inferred distance, so a
fractional distance offset $\epsilon = \delta d/d$ can generate a velocity
error of order $\epsilon\vc$.  For $\epsilon = 0.05$ and
$\vc \approx 200\kms$ this scale is $\sim10\kms$, but the actual bias in
$\vc$ also depends on geometry, selection, distance inference, and the fitted
spatial gradients.

\subsection{Non-equilibrium perturbations}

The Sagittarius dwarf galaxy's multiple pericentre passages excite bending
waves in the disc with amplitudes $\sim 1$--$2\kpc$ and associated
non-circular motions of $\sim 10$--$20\kms$ at $R \sim 20\kpc$
\citep{laporte2018}.  The LMC ($M_{\rm LMC} \approx 1$--$2 \times 10^{11}\Msun$)
creates a dynamical-friction wake and induces a reflex motion of the MW of
order tens of km\,s$^{-1}$ \citep{conroy2021,erkal2021}.  The warp of the
outer disc beyond $R \sim 12\kpc$ \citep{chen2019} also violates the planar
symmetry assumed in equation~\eqref{eq:jeans_R}.  The magnitude and
ubiquity of these disequilibrium effects, as well as the observational
signatures of the Gaia phase spiral and of the LMC-induced reflex motion
of the outer halo, are reviewed in detail by \citet{vasiliev2025}.

Two qualifications are in order.  First, the severity of the
equilibrium problem depends on the merger history of the galaxy.  The
last major merger experienced by the MW (Gaia--Sausage--Enceladus) dates
back to $\sim 10\,$Gyr ago, whereas a typical spiral of comparable mass
experienced its last major merger more recently, on average $\sim
6\,$Gyr ago \citep[see][and references therein]{jiao2023}.  The
equilibrium assumption is therefore \textit{less} problematic for the MW
disc than for many external spirals, and in particular than for M31.
Second, the degree to which equilibrium is questionable depends on the
radius and on the tracer.  The Sagittarius- and LMC-induced perturbations
of the disc at $R \lesssim 30\kpc$ are at the level of $\sim 10$--$20\kms$
(see above), while the assumption of a relaxed,
steady-state tracer population is far more doubtful for the halo stars,
globular clusters, streams, and dwarf satellites at $r \gtrsim 100\kpc$
that underpin the virial-mass estimates of
Section~\ref{sec:indep}: at those radii the LMC reflex motion alone is
comparable to the tracer velocity dispersion, orbital periods approach a
Hubble time, and a sizeable fraction of the satellites may be on first
infall.  Equilibrium considerations thus cut both ways in the
disc--halo tension discussed below, and do not by themselves favour the
higher-mass solutions.

\subsection{Statistical correlations}
\label{sec:correlations}

\citet{oman2024} showed that the correct likelihood for RC measurements uses
the full covariance matrix:
\begin{equation}
  \mathcal{L} \propto \exp\!\left(-\tfrac{1}{2}\bm{\Delta}^T\mathbf{C}^{-1}\bm{\Delta}\right),
  \label{eq:full_likelihood}
\end{equation}
where $\mathbf{C}_{ij} = \sigma_i\sigma_j r_{ij}$ includes off-diagonal
correlations between adjacent bins.  \citet{oman2024} found correlated
velocity fluctuations below about $10\kms$ over radial scales of
$1.5$--$2.5\kpc$ in simulated RCs; accounting for them can broaden inferred
halo parameters and shift total-mass estimates substantially.  Their result
does not by itself provide a recalibrated significance for the Jiao+23
Keplerian claim, but it shows that a diagonal likelihood can overstate the
effective number of independent radial constraints.

\section{Alternative Gravity Theories}
\label{sec:altgrav}

\subsection{MOND: framework and flat-RC prediction}

Modified Newtonian Dynamics \citep{milgrom1983} modifies the effective
gravitational force at accelerations below $a_0 \approx 1.2 \times
10^{-10}\,\mathrm{m\,s^{-2}}$:
\begin{equation}
  \mu\!\left(\frac{g}{a_0}\right) g = g_{\rm bar},
  \label{eq:mond}
\end{equation}
where $\mu(x) \to 1$ for $x \to \infty$ and $\mu(x) \to x$ for $x \to 0$.
In the deep-MOND regime ($g \ll a_0$), $\mu(g/a_0) \approx g/a_0$, so
equation~\eqref{eq:mond} becomes $g^2/a_0 = g_{\rm bar}$.  For circular
orbits ($g = \vc^2/R$):
\begin{equation}
  \vc^4 = GM_{\rm bar}\,a_0,
  \label{eq:btfr_mond}
\end{equation}
the baryonic Tully-Fisher relation (BTFR): the asymptotic circular speed
depends only on the total baryonic mass and $a_0$, predicting a flat RC
independent of radius.

\subsection{Tension between MOND and the declining MW RC}

At $R \sim20\kpc$ the MW is in the transition toward the low-acceleration
regime.  Approximating the outer baryonic field by the point-mass limit for
$M_{\rm bar}=(0.6$--$1.0)\times10^{11}\Msun$ gives
\begin{equation}
  g_{\rm bar}(R = 20\kpc)
  \approx \frac{G M_{\rm bar}}{(20\kpc)^2}
  \approx (0.17\text{--}0.29)\,a_0.
  \label{eq:g_bar_mw}
\end{equation}
Only farther out does the asymptotic result
equation~\eqref{eq:btfr_mond} predict a radius-independent speed
$\vc^{\rm MOND}=(GM_{\rm bar}a_0)^{1/4}\approx175$--$200\kms$.
A strict Keplerian decline is therefore difficult to accommodate in an
isolated standard-MOND model.  \citet{coquery2025} found that a standard
baryonic model with the standard MOND acceleration scale fails to reproduce
the declining RC.  When both the baryonic model and $a_0$ were allowed to
vary, the fit preferred a heavy stellar disc of order $10^{11}\Msun$ and an
$a_0$ consistent with zero, with a reported $95\%$ upper bound
$a_0<0.53\times10^{-10}\,\mathrm{m\,s^{-2}}$ rather than a value twice the
standard one.

Several caveats should be noted before regarding this as a clean
falsification of MOND.  First, the flat-RC prediction of
equation~\eqref{eq:btfr_mond} is strictly an \textit{asymptotic} result,
valid when $g_{\rm bar} \ll a_0$ in the outer field; at
$R \sim 20\kpc$ the MW is in the transition toward that regime, not
asymptotically deep within it, and corrections from the interpolating function
$\mu(x)$ can modify the predicted curve at the $\sim 10\%$ level.
Second, the MW is not an isolated system: the external-field effect
(EFE), a genuine prediction of MOND arising from its breaking of the
strong equivalence principle, can produce a declining outer RC for
galaxies embedded in the gravitational field of a larger structure
\citep{famaey2005}.  \citet{chae2020} have argued for statistical
evidence of the EFE in the SPARC sample (Spitzer Photometry and Accurate
Rotation Curves; \citealt{lelli2016}), although this interpretation
is contested.  Whether a plausible external field (e.g.\ from the
Local Group or from large-scale structure) can reproduce the observed
MW decline quantitatively is an open question and has not, to our
knowledge, been assessed in detail against the Jiao+23 data.  Third,
hybrid DM/MOND frameworks (e.g.\ neutrino hot dark matter or a light
sterile component in a MOND potential) can in principle soak up the
residual at outer radii without abandoning the BTFR for isolated
galaxies.  We therefore regard MOND as being in tension with the
strict Keplerian interpretation of the MW RC, but the inference remains
conditional on the baryonic model, equilibrium RC measurement, interpolating
function, and external field.

\begin{figure}[htbp]
\centering
\includegraphics[width=\linewidth,keepaspectratio]{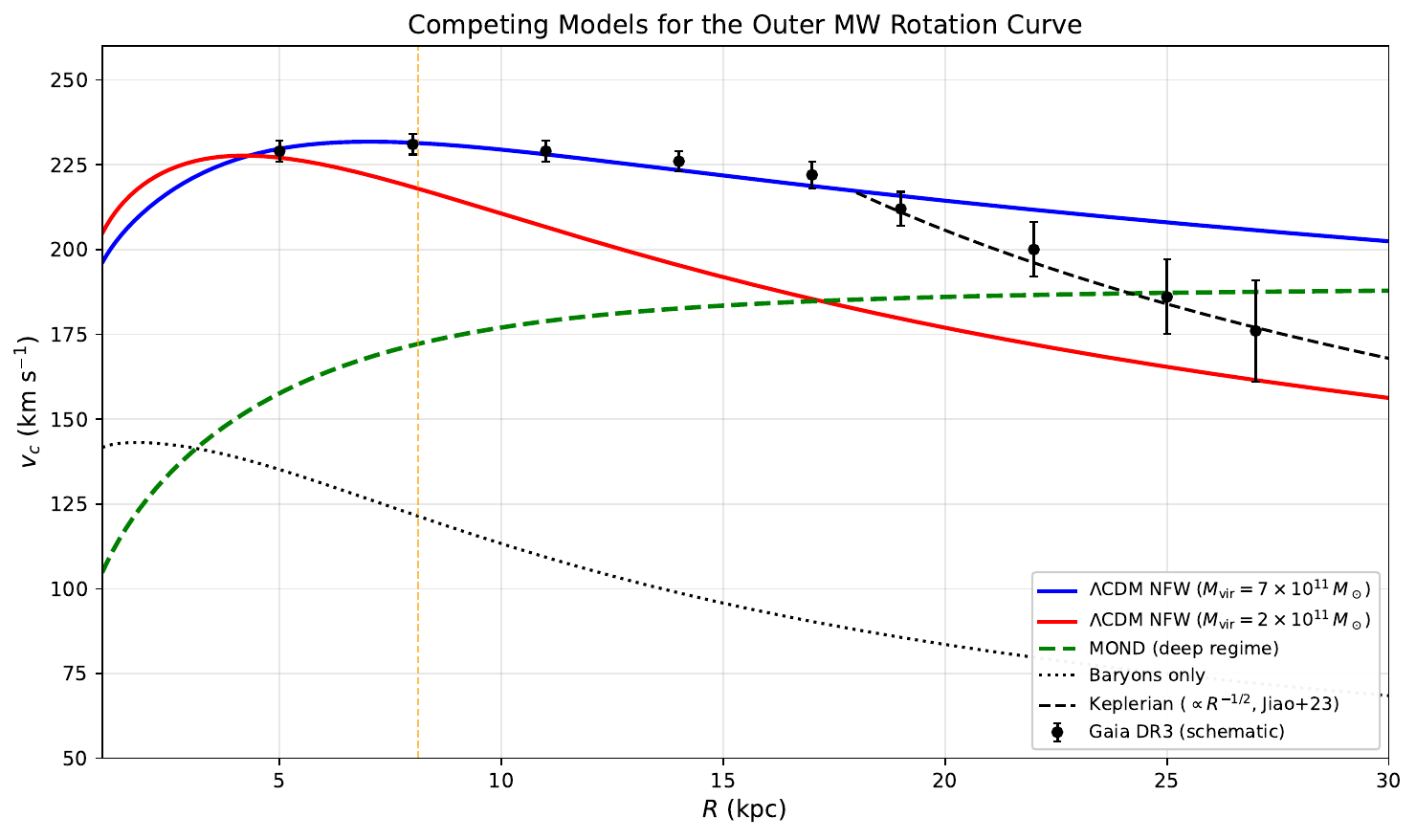}
\caption{Competing theoretical models for the outer MW RC.  The high-mass
  NFW (blue solid) gives a nearly flat RC.  The low-mass NFW (red solid) has
  a steeper decline.  The MOND prediction (green dashed) asymptotes to a flat
  $\approx 175$--$200\kms$ asymptotically.  The Keplerian
  decline from $R = 19\kpc$ (black dashed) and the schematic Gaia DR3 data
  (grey circles) are shown for comparison.}
\label{fig:models_compare}
\end{figure}

Figure~\ref{fig:models_compare} shows the qualitative differences among
representative Newtonian-halo, Keplerian, and MOND curves.

\subsection{Modified Gravity (MOG)}

The MOG theory \citep{moffat2006} introduces an enhanced gravitational
coupling $G = G_N(1+\alpha_{\rm MOG})$, partially screened at short range
by a repulsive Yukawa contribution.  The point-mass potential is:
\begin{equation}
  \Phi_{\rm MOG}(r)
  = -\frac{G_N M}{r}
    \bigl(1 + \alpha_{\rm MOG} - \alpha_{\rm MOG}\,e^{-r/\lambda}\bigr),
  \label{eq:mog_phi}
\end{equation}
which correctly reduces to the Newtonian potential $-G_N M/r$ both for
$\alpha_{\rm MOG} \to 0$ and for $r \ll \lambda$, and asymptotes to
$-G_N(1+\alpha_{\rm MOG})M/r$ for $r \gg \lambda$.

\citet{moffat2025} fit the \citeauthor{jiao2023} data with $\alpha_{\rm MOG}
= 13$ and $\lambda = 25\kpc$, obtaining $M_{\rm tot} \approx 1.3 \times
10^{11}\Msun$, attributing the residual mass to the hot plasma of the
circumgalactic medium (CGM).

\subsection{General relativistic approaches}

\citet{beordo2024} compared MOND, dark-matter halo, and general-relativistic
models with Gaia-based data limited to $R\lesssim19\kpc$ and found them
statistically comparable over that interval.  Those data do not reach the
claimed Keplerian outer tail, so the comparison cannot adjudicate its
existence; the paper does not establish a detected or predicted
$\sim10\kms$ frame-dragging signature in the outer-disc RC.

\section{Independent Dynamical Constraints}
\label{sec:indep}

\subsection{Stellar streams}

Stars stripped from a satellite near the Roche radius (equation~\ref{eq:roche}
below) form tidal streams that trace orbits in the MW potential.  The
Orphan-Chenab stream \citep{koposov2023} extends to $r \sim 60\kpc$,
constraining:
\begin{equation}
  M_{\rm MW}(< 50\kpc) =
  3.77^{+0.26}_{-0.19} \times 10^{11}\Msun,
\end{equation}
already exceeding the total mass from the Keplerian interpretation
(equation~\ref{eq:jiao_mass}).  The Sagittarius stream gives
$M(<100\kpc)=5.6\pm0.4\times10^{11}\Msun$ and
$\Mvir=9.0\pm1.3\times10^{11}\Msun$ in the adopted halo family
\citep{vasiliev2021}.

For a satellite on a circular orbit, in the point-mass/local tidal
approximation, the characteristic Roche radius is
\begin{equation}
  r_{\rm Roche}\simeq
  = r_{\rm orb}
    \left(\frac{M_{\rm sat}}{3M_{\rm MW}(r_{\rm orb})}\right)^{1/3}.
  \label{eq:roche}
\end{equation}

\subsection{Globular cluster kinematics}

Applying the spherical Jeans equation~\eqref{eq:jeans_sph_sol} to samples of
order $10^2$ globular clusters with line-of-sight velocities and, for
subsamples, proper motions:
\begin{equation}
  \nu_{\rm GC}(r)\sigma_r^2(r)
  = r^{-2\beta_0}
    \int_r^\infty s^{2\beta_0}\,\nu_{\rm GC}(s)\,\frac{GM(s)}{s^2}\,\dif s.
  \label{eq:gc_jeans}
\end{equation}
\citet{eadie2019} obtained a median
$M_{200}=0.70\times10^{12}\Msun$ with a $50\%$ interval
$0.62$--$0.81\times10^{12}\Msun$, and
$M(<100\kpc)=0.53^{+0.21}_{-0.12}\times10^{12}\Msun$.
\citet{posti2019}: $\Mvir = (1.3 \pm 0.3) \times 10^{12}\Msun$.

\subsection{The Local Group timing argument}

The MW--M31 system is modelled as a two-body problem with parametric
Keplerian solution \citep{kahn1959}:
\begin{align}
  r(\eta) &= A(1 - \cos\eta),\quad
  t(\eta) = B(\eta - \sin\eta),\quad A^3 = GM_{\rm LG}\,B^2.
  \label{eq:timing}
\end{align}
Given $(r_0 = 770\kpc$, $v_r = -110\kms$, $t_0 = 13.8\,\mathrm{Gyr})$,
numerical solution gives $M_{\rm LG} \approx (4.7 \pm 0.6) \times
10^{12}\Msun$ \citep{benisty2024}.  This constrains the combined MW--M31
mass and depends on the treatment of cosmic expansion, tangential motion,
the LMC, and the mass split between the two galaxies; it does not by itself
imply a model-independent lower bound $M_{\rm MW}\gtrsim10^{12}\Msun$.
A complementary dynamical analysis of the M31 rotation curve by
\citet{hammer2025}, using the same methodology that \citet{jiao2023}
applied to the MW, derives a low M31 dynamical mass with baryon fraction
$\sim 30\%$ within the virial radius.  Taken together with the Jiao+23
MW result, this would push the Local Group total mass towards the lower
end of the timing-argument range; however, \citeauthor{hammer2025} explicitly
note that a Keplerian decline cannot be detected in M31 because of its
strongly perturbed post-merger kinematics, so this M31 mass estimate is
not a clean external check on the MW decline.

\subsection{Virial mass compilation}

A recent independent constraint comes from \citet{hayati2025}, who use
a neural-network approach to infer the MW halo mass from the dynamics
of its satellite system without assuming dynamical equilibrium or that
all neighbouring galaxies are bound satellites.  They obtain
$\log_{10}(\Mvir/\Msun) = 12.20^{+0.16}_{-0.14}$, i.e.\ $\Mvir \approx
1.6\times10^{12}\Msun$, in good agreement with the stream, GC, and
other satellite-based estimates and in strong tension with the Jiao+23
Keplerian value.

Table~\ref{tab:mass_comp} compares these heterogeneous constraints while
distinguishing enclosed masses from virial masses.

\begin{table}[htbp]
\centering
\caption{MW mass constraints from different methods.  All masses are in units
  of $10^{11}\Msun$; $M_{200}$ or $M_{\rm vir}$ is shown unless an enclosed
  radius is stated explicitly.  GC: globular clusters; NN: neural network.}
\label{tab:mass_comp}
\scriptsize
\begin{tabular*}{\linewidth}{@{\extracolsep{\fill}}lccc}
\toprule
Method & Study & Mass constraint & Notes\\
\midrule
RC (NFW)        & Eilers+19     & $7.25\pm0.26$          & to 25 kpc   \\
RC (NFW)        & Sylos L.+23   & $6.5\pm0.5$            & to 28 kpc   \\
RC (gNFW)       & Ou+24         & $M_{200}=6.94\pm0.12$  & poor fit    \\
RC (Einasto)    & Ou+24         & $M_{200}=1.81^{+0.06}_{-0.05}$ & preferred fit \\
RC (Keplerian)  & Jiao+23       & $M_{\rm tot}=2.06^{+0.24}_{-0.13}$ & spherical interpretation \\
Stream          & Koposov+23    & $M(<50)=3.77^{+0.26}_{-0.19}$ & Orphan--Chenab \\
Stream          & Vasiliev+21   & $M_{\rm vir}=9.0\pm1.3$ & Sagittarius \\
GC              & Posti+19      & $M_{\rm vir}=13\pm3$   & 75 GCs      \\
GC              & Eadie+19      & $M_{200}=7.0^{+1.1}_{-0.8}$ & 50\% interval \\
Satellites      & Cautun+20     & $10.8^{+2.0}_{-1.4}$  &             \\
Satellites (NN) & Hayati+25     & $16^{+7}_{-4}$         & neural net, no equilibrium assumed \\
Timing arg.     & Benisty+24    & $M_{\rm LG}=47\pm6$    & MW+M31      \\
\bottomrule
\end{tabular*}
\end{table}

\section{Dark Matter Density and Particle Physics Implications}
\label{sec:dm}

\subsection{The local dark matter density}

The expected event spectrum in a direct-detection experiment for weakly
interacting massive particles (WIMPs) scales linearly with
$\rholocDM$ (equation~\ref{eq:dd_rate_full} in \ref{app:units}).
For the canonical NFW model with $r_s = 14.8\kpc$ and
$\rho_s = 1.06 \times 10^7\Msun\kpc^{-3}$, using equation~\eqref{eq:nfw_rho}:
\begin{align}
  \rholocDM
  &= \frac{\rho_s}{(R_0/r_s)(1 + R_0/r_s)^2}
   = \frac{1.06 \times 10^7}{0.553 \times 2.41}
   \approx 0.30\,\mathrm{GeV\,cm^{-3}}.
  \label{eq:rho_local}
\end{align}
The local density cannot be inferred from a virial mass alone.  Indeed, the
two \citet{ou2024} halo fits have very different $M_{200}$ values but give
$\rholocDM=0.447\pm0.004\,\mathrm{GeV\,cm^{-3}}$ (Einasto) and
$0.405\pm0.004\,\mathrm{GeV\,cm^{-3}}$ (gNFW).  A self-consistent local
constraint must combine the inner RC or vertical force with a specified
baryonic and halo geometry; an order-of-magnitude reduction does not follow
from the Jiao+23 total-mass estimate.

\subsection{The escape speed}
\label{sec:vesc}

Using the NFW potential of equation~\eqref{eq:nfw_phi} and the standard
relation $v_{\rm esc}^2 = 2|\Phi|$ (see \ref{app:nfw_phi}):
\begin{equation}
  v_{\rm esc,NFW}^2(r)
  = 2|\Phi_{\mathrm{NFW}}(r)|
  = \frac{8\pi G\rho_s r_s^3}{r}\ln\!\left(1+\frac{r}{r_s}\right).
  \label{eq:vesc_nfw}
\end{equation}
For the canonical NFW halo ($r_s = 14.8\kpc$,
$\rho_s = 1.06\times 10^7\Msun\kpc^{-3}$), equation~\eqref{eq:vesc_nfw}
gives a \textit{halo-only} escape speed of $v_{\rm esc}(R_0) \approx
447\kms$; adding the baryonic contribution to the potential
($M_{\rm bar} \approx 7\times 10^{10}\Msun$) raises the total to
$v_{\rm esc}(R_0) \approx 530$--$550\kms$.  Any low-mass alternative must
be tested with its full baryonic and halo potential rather than by rescaling
$M_{200}$ alone.  Gaia measurements of high-velocity stars
\citep{deason2019} give $v_{\rm esc}(R_0) = 528^{+24}_{-25}\kms$, providing
an important complementary constraint.

\subsection{Implications for microlensing}

The optical depth toward the LMC:
\begin{equation}
  \tau_{\rm lens}
  = \frac{4\pi G}{c^2}\int_0^{D_s}
    \rho_{\rm DM}(D_l)\,\frac{D_l(D_s-D_l)}{D_s}\,\dif D_l.
  \label{eq:tau_lens}
\end{equation}
Note that $\tau_{\rm lens}$ depends only on the mass \textit{density} along
the line of sight and not on the individual lens mass: the lens number
density and the Einstein-radius cross-section compensate exactly.  The
lens mass instead enters through the event durations and rate.
\citet{garciabellido2024} found that adopting the \citeauthor{jiao2023} density
profile allows a fraction of massive compact halo objects (MACHOs) up to
$20\%$ for masses
$10^{-3}$--$1\Msun$---relaxing the earlier tight constraints.

\section{The Radial Acceleration Relation}
\label{sec:rar}

\subsection{Definition and observational status}

The RAR \citep{mcgaugh2016} correlates observed centripetal acceleration
$g_{\rm obs} = \vc^2/R$ with the baryonic prediction
$g_{\rm bar} = v_{c,\rm bar}^2/R$:
\begin{equation}
  g_{\rm obs}
  = \frac{g_{\rm bar}}{1 - \exp\!\bigl(-\sqrt{g_{\rm bar}/g_\dagger}\bigr)},
  \quad g_\dagger \approx 1.2 \times 10^{-10}\,\mathrm{m\,s^{-2}},
  \label{eq:rar}
\end{equation}
holding for 2693 resolved points in 153 SPARC galaxies, with reported scatter
of order $0.13\,\mathrm{dex}$.  The universality of the relation has been
questioned: \citet{dipaolo2019} find that dwarf disc and low-surface-brightness
galaxies deviate systematically from equation~\eqref{eq:rar}, with a
dependence on galaxy radius and luminosity that points to an intrinsic
scatter larger than reported for the SPARC sample, so that the RAR may be a
projection of the underlying DM--baryon scaling relations rather than a
fundamental law.

\subsection{The MW on the RAR}

At $R=20\kpc$, the approximate range in
equation~\eqref{eq:g_bar_mw}, $g_{\rm bar}\simeq(0.17$--$0.29)a_0$, gives
an RAR speed of roughly $194$--$227\kms$.  This brackets the
$\sim207$--$212\kms$ values represented by the Jiao+23 and Ou+24 RCs at
that radius.  The comparison is therefore sensitive to the adopted baryonic
force; using a substantially smaller baryonic contribution would instead
produce an apparent offset.  A meaningful placement of the MW on the RAR
requires a three-dimensional bulge, disc, and gas model with propagated
uncertainties; the crude point-mass calculation neither establishes an
RAR anomaly nor removes the separate question of the outer RC slope.

\begin{figure}[htbp]
\centering
\includegraphics[width=0.85\linewidth,keepaspectratio]{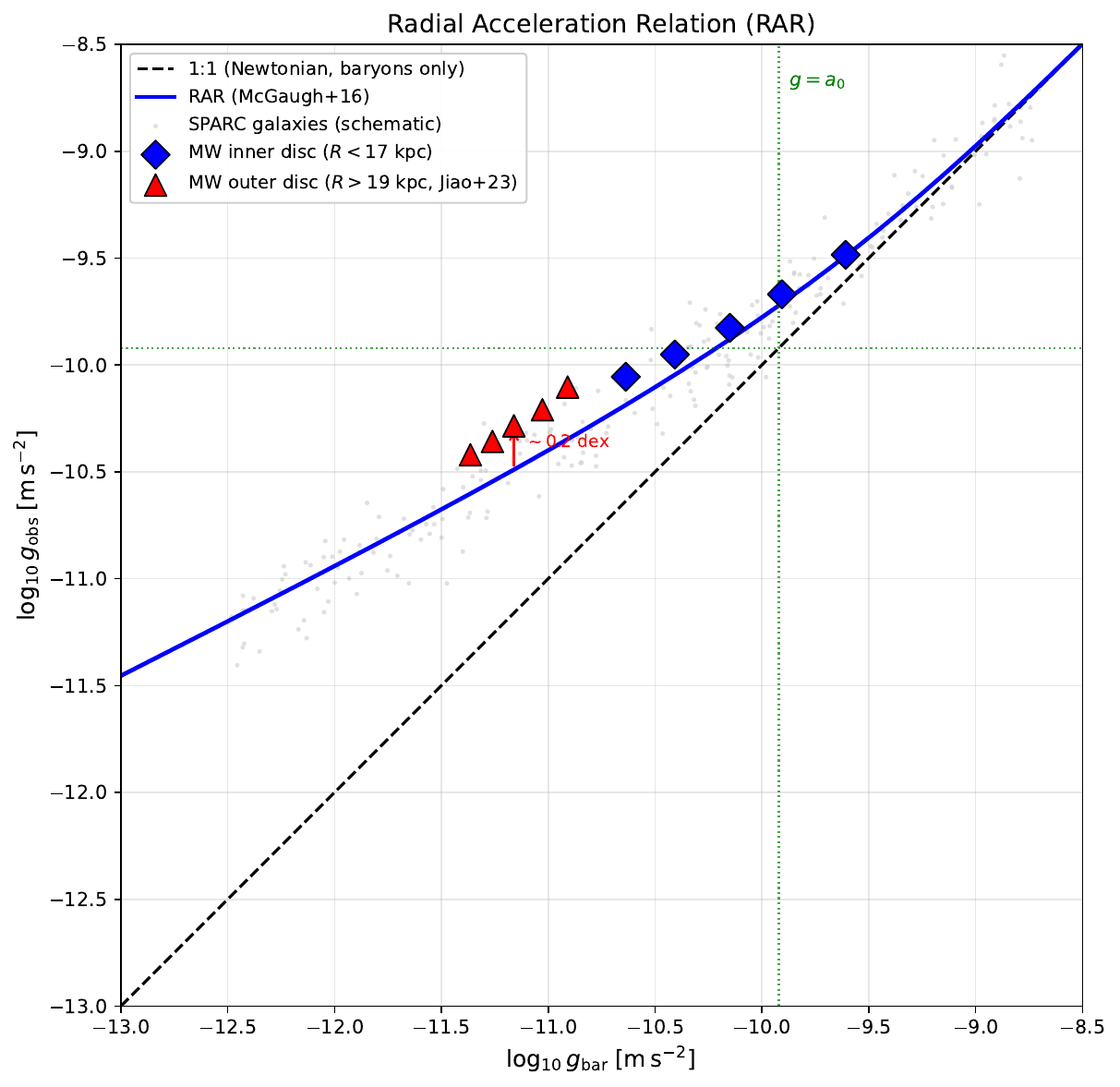}
\caption{The Radial Acceleration Relation.  Grey points: SPARC galaxies
  \citep{lelli2016}.  Blue solid: best-fit RAR (equation~\ref{eq:rar},
  \citealt{mcgaugh2016}).  Black dashed: 1:1 Newtonian line.  Blue diamonds:
  MW inner disc ($R < 17\kpc$).  Red triangles: MW outer disc under the
  Keplerian interpretation (\citealt{jiao2023}); their placement depends on
  the adopted MW baryonic model.
  Green dotted lines mark $a_0$.}
\label{fig:rar}
\end{figure}

Figure~\ref{fig:rar} illustrates the RAR comparison, with the MW location
understood to depend on the adopted baryonic model.

\section{Discussion: Reconciling the Observations}
\label{sec:discussion}

\subsection{Three possible scenarios}

Three internally consistent scenarios can accommodate the full set of
observations.

In \textit{Scenario 1}, the Keplerian decline is real and the outer-halo mass
tracers are systematically biased upward or are being compared across
incompatible halo models.  This would require multiple
independent methods (streams, GCs, satellites, timing argument) to fail
simultaneously in the same direction---unlikely but not logically excluded.

\textit{Scenario 2} is a plausible intermediate interpretation: there is a genuine decline at
$R > 15\kpc$, but its magnitude is amplified by systematic errors in the
Jeans methodology.  The true virial mass lies in the range
$(5$--$8) \times 10^{11}\Msun$, consistent with the lower end of stream-based
estimates.  The FIRE-2 analysis of \citet{ou2024_fire} supports this
interpretation, finding biases that can reach $\sim10$--$40\kms$ at
$R\sim20$--$25\kpc$ in simulated observational pipelines.

In \textit{Scenario 3}, the decline is entirely an artefact of the
cylindrical Jeans methodology, and the true RC is nearly flat at
$\vc \approx 220$--$230\kms$ out to $R \sim 30\kpc$, consistent with
$\Mvir \approx 10^{12}\Msun$.  This scenario draws support from the
concerns about tracer geometry and closure assumptions in the
halo-dominated regime discussed in
Section~\ref{sec:jeans_inconsistency} (cf.\ \citealt{klacka2025})
and by the fact that no external galaxy of comparable luminosity shows a
Keplerian decline at equivalent radii: the universal rotation curve of
spirals of MW luminosity is flat or gently declining out to the last
measured point \citep{persic1996,salucci2019,lelli2016,mistele2024,bariego2026}.

\subsection{The puzzle of MW exceptionalism}

If the decline is genuine, it raises the question: why is the MW different
from the predominantly flat or gently declining RCs of comparable spiral
galaxies observed in 21\,cm HI, which do not show an ensemble Keplerian
taper at equivalent radii?  Three possible reconciliations are: (i) the observational
methods are fundamentally different (21\,cm vs.\ stellar Jeans), with
different systematic floors; (ii) the MW has an unusually concentrated halo
($c \sim 15$--$20$); or (iii) the outer disc is genuinely perturbed by the
Sgr and LMC interactions, and its apparent kinematics do not reflect the
equilibrium circular speed.

A dedicated statistical test of this apparent exceptionalism was carried
out by \citet{bariego2026}, who applied a uniform
slope-fitting procedure to the DM-subtracted rotation curves of all 175
SPARC galaxies \citep{lelli2016}, testing whether the terminal slope
$V'(r)$ is compatible with the flat (Rubin) hypothesis or with a Keplerian
taper.  Using parametric ($Z$) and non-parametric (sign, Wilcoxon) tests,
they find that the SPARC sample is fully consistent with a null terminal
slope --- the binomial, Wilcoxon, and $Z$-test $p$-values all cluster
around $0.9$--$1.0$, providing no statistical evidence for a Keplerian
decline in external spirals.  Applying the same pipeline to the modern MW
data of \citet{eilers2019}, \citet{wang2023}, \citet{zhou2023}, \citet{jiao2023},
and \citet{ou2024} yields, by contrast, a clearly negative terminal slope
of the DM contribution of $-1.57 \pm 0.11\kms\kpc^{-1}$ in the combined
sample for $R > 15\kpc$.  The contrast between the MW and the SPARC
sample is therefore sharpened, rather than resolved, by a homogeneous
analysis: whatever is driving the apparent decline in the MW does not
appear to be operating (or is not yet detectable) in the external
spirals at comparable fractional radii.  The authors suggest that
deeper radio surveys such as MeerKAT's MHONGOOSE and the future SKA
may be needed to push SPARC-like rotation curves to the distances where
a decline, if present, would become statistically visible.

\subsection{Constraints on dark matter physics}

Even setting aside the Keplerian vs.\ sub-Keplerian ambiguity, the declining
RC constrains the DM halo profile shape.  The \citet{ou2024} gNFW fit has
$\gamma\simeq0.026$, but its poor reduced $\chi^2$ makes a quantitative
claim for a core model-dependent; their preferred Einasto fit instead has
an unusually large shape parameter $\alpha\simeq0.91$.  Such inferences are
sensitive to baryonic feedback \citep{pontzen2014}, possible DM
self-interactions, and the RC systematics.  Warm DM is not favoured because current
Lyman-$\alpha$ and substructure lensing constraints disfavour the thermal
relic masses needed to produce truncation at $\sim 20\kpc$.

An even sharper tension with $\Lambda$CDM is obtained by
\citet{li2025}, who reverse-engineer the \textit{primordial} (pre-baryonic
compression) DM halo of the MW by iteratively undoing the contraction
induced by the observed baryonic distribution, using the Jiao+23 RC as
input.  The inferred primordial profile differs from CDM predictions by
more than $3\sigma$ in both concentration and total mass, and combines
two features that are individually problematic and jointly unnatural:
a shallow inner core (harder to generate in a massive halo than in
dwarfs because baryonic compression dominates over supernova feedback)
and an unusually steep outer decline.  If the declining RC is genuine,
these features require either non-trivial DM microphysics (e.g.\ SIDM or
fuzzy DM) combined with unusually strong baryonic feedback, or a
substantial revision of current galaxy-formation models.  This result
is therefore tightly coupled to whether the outer decline survives the
systematic tests discussed in Section~\ref{sec:systematics}: a
moderate decline is comfortably accommodated by NFW/gNFW fits, but a
genuine Keplerian tail is difficult to reconcile with any CDM-based
halo model.

\section{Conclusions}
\label{sec:conclusions}

The rotation curve of the Milky Way has been transformed by \textit{Gaia}
from one of the most uncertain to one of the most precisely measured in any
spiral galaxy.  Our central conclusions are as follows.

Several Gaia-era analyses find a decline beyond $R\gtrsim15\kpc$, but its
magnitude and even the robustness of the outermost tail remain contested
because the analyses share distance calibrations, tracer assumptions, and
equilibrium modelling.  Cepheid measurements to $18$--$20\kpc$ favour only
a mild decline, whereas RGB-based analyses extending to $25$--$28\kpc$
often find a steeper one.  Whether the \textit{true} circular-speed curve
declines as steeply as the Jeans-inferred curve therefore remains open
(Scenario~3 of Section~\ref{sec:discussion}).

The Keplerian interpretation is plausible but systematically challenged.  The
claim of \citet{jiao2023} that the decline is Keplerian at $3\sigma$ rests on
steps---distance reconciliation, the assumed tracer density and dispersion
gradients (which enter both the asymmetric drift correction and the radial
pressure-support term in a correlated way), and a simplified axisymmetric
Jeans analysis---can generate correlated biases when propagated through the
full inference pipeline.  FIRE-2 simulations find combined biases of
$\sim 10$--$40\kms$ at $R > 20\kpc$
\citep{ou2024_fire}.  In addition, at these radii the gravitational
acceleration is dominated by the (nearly spherical) halo, a regime in which
the tracer-geometry and closure assumptions underlying the cylindrical
Jeans analysis are least well tested \citep{klacka2025}.

Off-plane information provides a complementary test of geometry.
\citet{syloslabini_capuzzo2026} find that a massive, co-rotating DMD fits
their reconstructed $v_c(R,z)$ and $a_z(R,z)$ inside $14\kpc$ better than a
spherical NFW halo.  This is an important proof of concept, but it neither
samples the claimed Keplerian tail nor yet provides a model-independent
detection of a dark disc: its preference depends on an effective
multicomponent closure, model-conditioned vertical scale heights, omission
of the tilt term, and comparison of only two limiting geometries.

RC-based virial masses are strongly profile-dependent: for example, the two
\citet{ou2024} fits span $M_{200}\simeq1.8$--$6.9\times10^{11}\Msun$ despite
using the same RC.  Many stream, globular-cluster, and satellite analyses
favour roughly $(0.7$--$1.6)\times10^{12}\Msun$, with substantial modelling
dependence.  The Jiao+23 spherical-equivalent total mass of
$2 \times 10^{11}\Msun$ is in strong tension with those outer-halo
constraints.

An isolated MOND model predicts an asymptotically flat RC, so a confirmed
Keplerian tail would be difficult to accommodate.  The MW at $20\kpc$ is
still in the transition toward the deep-MOND regime, however, and the
quantitative test depends on the baryonic model, interpolating function, and
external field.  \citet{coquery2025} found that relaxing the baryonic model
drives the fitted $a_0$ below, not above, the standard value.

The local DM density is not determined by the virial mass alone.  It must be
inferred from the inner RC or vertical force in a joint baryonic--halo model;
the Ou+24 fits illustrate that very different virial masses can yield similar
local densities.

Resolution requires better data and improved modelling: \textit{Gaia} DR4,
scheduled for 2 December 2026,\footnote{\url{https://www.cosmos.esa.int/web/gaia/release}}
WEAVE, 4MOST, and DESI for kinematics; SKA-MID for maser parallaxes;
and a consistent three-dimensional Jeans analysis validated against
cosmological simulations, using the spherical (or oblate) Jeans equation at
large radii.

\section*{Acknowledgements}

The authors thank the Sapienza cosmology group and INFN Roma1 for discussions
and support.  We are grateful to Fran\c{c}ois Hammer for detailed and
constructive comments on the first arXiv version of this review, in
particular concerning the placement of the outer \citet{eilers2019} data
points in Figure~\ref{fig:rc_data}, the orbital properties of the outer-disc
tracers of \citet{ou2024}, and the applicability of equilibrium assumptions
to the Milky Way disc versus its outer halo.  We also thank Yang Huang for
comments on the distance scale of \citet{zhou2023}, on the statistical
nature of the \citet{wang2023} distances, and on the limited leverage of
the disc rotation curve on the total Galactic mass.  This work was supported by INFN Roma1 and by the Italian
Ministry of University and Research.  R.\ also acknowledges support from
grant PID2024-158938NB-I00, funded by MICIU/AEI/10.13039/501100011033 and by
``ERDF A way of making Europe'', and from Project SA097P24, funded by the
Junta de Castilla y Le\'on.

\section*{Data and Code Availability}

The rotation-curve measurements used in the phenomenological analysis are
listed in Table~\ref{tab:rc_data}, with their original sources identified in
its caption.  The numerical code, parameter choices, Markov chains,
convergence outputs, and scripts used to produce the MCMC and Gaussian
Process figures are available from the authors upon reasonable request.

\appendix

\section{Useful Unit Conversions}
\label{app:units}

\begin{table}[h]
\centering
\caption{Frequently used unit conversions.}
\begin{tabular}{lll}
\toprule
\textbf{Quantity} & \textbf{Relation} & \textbf{Value}\\
\midrule
Grav.\ constant  &$G$&$4.302\times10^{-6}\,\mathrm{kpc}\,\Msun^{-1}(\kms)^2$\\
Acceleration &$1\,(\kms)^2\kpc^{-1}$&$=3.241\times10^{-14}\,\mathrm{m\,s^{-2}}$\\
DM density   &$1\,\Msun\,\mathrm{pc}^{-3}$&$=37.96\,\mathrm{GeV\,cm^{-3}}$\\
Solar radius &$\Rsun$&$=8.178\kpc$\\
Orbital period&$T=2\pi\Rsun/\vc(R_0)$&$\approx220\,\mathrm{Myr}$\\
MOND $a_0$  &$1.2\times10^{-10}\,\mathrm{m\,s^{-2}}$&$=3.70\times10^{3}\,(\kms)^2\kpc^{-1}$\\
Crit.\ density&$\rho_{\rm crit}$&$=1.26\times10^{11}\Msun\,\mathrm{Mpc}^{-3}$\\
\bottomrule
\end{tabular}
\end{table}

The exposure-integrated event spectrum referenced in the text:
\begin{equation}
  \frac{\dif N}{\dif E_R}
  = \frac{M_T\tau\,\rholocDM}{m_\chi m_N}
    \int_{v > v_{\rm min}}\! v\,f(\bm{v})\,
    \frac{\dif\sigma_{\chi N}}{\dif E_R}\,\dif^3\bm{v}.
  \label{eq:dd_rate_full}
\end{equation}
Here $E_R$ is the nuclear recoil energy, $M_T$ the detector target mass,
$\tau$ the exposure time, $m_\chi$ and $m_N$ the DM-particle and target-nucleus
masses, $f(\bm{v})$ the local DM velocity distribution in the detector
frame, $v_{\rm min}$ the minimum speed able to produce a recoil of energy
$E_R$, and $\dif\sigma_{\chi N}/\dif E_R$ the differential
DM--nucleus cross-section.

\section{General Solution of the Spherical Jeans Equation}
\label{app:jeans_sph}

For constant anisotropy $\beta_0$, multiplying equation~\eqref{eq:jeans_sph}
by $r^{2\beta_0}$ gives:
\begin{equation}
  \frac{\dif}{\dif r}\bigl[r^{2\beta_0}\nu\sigma_r^2\bigr]
  = -r^{2\beta_0}\nu(r)\frac{GM(r)}{r^2}.
\end{equation}
Integrating from $r$ to $\infty$ recovers equation~\eqref{eq:jeans_sph_sol}
directly.  For $\beta_0 = 0$: $\nu\sigma_r^2 = \int_r^\infty
\nu(s)GM(s)/s^2\,\dif s$.

\section{NFW Gravitational Potential Derivation}
\label{app:nfw_phi}

Integrating $\dif\Phi/\dif r = GM_{\rm NFW}(r)/r^2$ with
$\Phi(\infty)=0$
using the mass from equation~\eqref{eq:nfw_mass}, one obtains
equation~\eqref{eq:nfw_phi}.  The escape speed:
\begin{equation}
  v_{\rm esc}^2(r)
  = 2|\Phi_{\mathrm{NFW}}(r)|
  = \frac{8\pi G\rho_s r_s^3}{r}\ln\!\left(1+\frac{r}{r_s}\right).
  \label{eq:vesc_nfw_app}
\end{equation}

\section{The Baryonic Tully-Fisher Relation}
\label{app:btfr}

From equation~\eqref{eq:btfr_mond}, the BTFR can be written as:
\begin{equation}
  M_{\rm bar}
  = \frac{1}{Ga_0}\vc^4
  \simeq 62.8\,\Msun\,(\kms)^{-4} \times \vc^4,
\end{equation}
using $G = 4.302\times10^{-6}\kpc\,\Msun^{-1}(\kms)^2$ and
$a_0 = 3.70\times10^3\,(\kms)^2\kpc^{-1}$.
For the MW with $M_{\rm bar} \approx 7\times10^{10}\Msun$ and
$\vc(R_0) \approx 229\kms$, the dimensionless ratio is
$G a_0 M_{\rm bar}/\vc^4(R_0) \approx 0.4$; equivalently, the deep-MOND
(BTFR) prediction for the asymptotic flat velocity is
$(GM_{\rm bar}a_0)^{1/4} \approx 183\kms$, below the observed
$\vc(R_0) \approx 229\kms$.  Exact equality is not expected at the solar
radius, where $g(R_0) = \vc^2/R_0 \approx 1.7\,a_0$ places the MW in the
transition (rather than deep-MOND) regime; in Newtonian terms, the excess
of $\vc(R_0)$ over the baryonic prediction quantifies the DM contribution
already required at $R_0$.

\bibliographystyle{elsarticle-harv}

\end{document}